\documentclass[aps,floatfix,showpacs]{revtex4}
\usepackage{amsmath}
\usepackage{bbm}
\usepackage{graphicx}

\begin{document}

\flushbottom

%\topical{Title}
\title{Superconducting nanostructures fabricated with the STM}

\author{J. G. Rodrigo, H. Suderow and S. Vieira\footnote[1]{Corresponding 
author: S. Vieira \\E-mail: sebastian.vieira@uam.es}}
\affiliation {Laboratorio de Bajas Temperaturas, Depto. de F\'isica de la Materia
Condensada, Instituto de Ciencia de Materiales Nicol\'as Cabrera,
Universidad Aut\'onoma de Madrid, 28049 Madrid, Spain. }
\author{E. Bascones}
\affiliation{ Theoretische Physik, ETH-H\"onggerberg, CH-8093 Z\"urich, Switzerland }
\author {F. Guinea}
\affiliation{Instituto de Ciencia de Materiales de Madrid. CSIC.
Cantoblanco. 28049 Madrid. Spain}

\date{\today}

\begin{abstract}
The properties of nanoscopic superconducting structures fabricated
with a scanning tunnelling microscope are reviewed, with emphasis
on the effects of high magnetic fields. These systems include the
smallest superconducting junctions which can be fabricated, and
they are a unique laboratory where to study superconductivity
under extreme conditions. The review covers a variety of recent
experimental results on these systems, highlighting their unusual
transport properties, and theoretical models developed for their
understanding.
\end{abstract}


\maketitle

\tableofcontents

\section{Introduction}
Soon after the transcendental discovery by Meissner of the perfect
diamagnetism\cite{Meissner}, as one of the characteristic features of the
superconducting state, the London brothers published in 1935 an
article entitled "The electromagnetic equations of the
superconductor"\cite{London35}. This article contains the well
known London theory, which provided the first important approach to
our macroscopic understanding of this phenomenon. Very soon one of
the brothers, Heinz London, concluded from the theory that "a very
small superconductor should have a much higher magnetic threshold
value than a bulky one"\cite{Londonh35}. R.B. Pontius 
confirmed this prediction experimentally two years
later\cite{Pontius,Pontius2}. In 1939 E.T.S. Appleyard et al.
\cite{Appleyard} found an increase of the magnetic threshold up to
more than twenty times the bulk critical field in mercury films as
thin as 57nm. D. Shoenberg\cite{Shoenberg} observed
this effect by measuring the magnetic susceptibility of very
fine-grained preparations of colloidal mercury in 1940. From then to now,
there have been many important developments both in theory and
experiment, including several milestones such as the Ginzburg
-Landau (G-L) theory\cite{GL50}, the microscopic BCS
theory\cite{BCS57}, the Josephson effects\cite{J62}, the 
type II superconductors\cite{A57} and the discovery of
high T$_c$ superconducting
oxides\cite{Bednorz85}. G-L theory has proven itself as a very
important tool which separates superconductors into two types (I
and II), depending on their response to external magnetic fields.
Abrikosov predicted\cite{A57} the real existence of
type II superconductors, with the characteristic mixed or vortex
state present in a wide range of magnetic fields. 
Tinkham\cite{Tinkham63} pointed out that sufficiently thin films of 
any material should exist in the mixed state even if thicker specimens 
of the same material exhibit a type I behavior.
This vortex state was visualized later on in magnetic decoration
experiments\cite{Essman67,Dolan73}.

Today superconductivity is one of the most flourishing fields of
condensed matter physics, showing many new interesting
developments. A recent one is the
reduction of the dimensions of the superconducting samples towards
controlled three dimensional mesoscopic structures. Electron lithography 
allows to pattern
different types of superconducting 
structures with all their dimensions of the same order or smaller 
than the magnetic
penetration depth of the bulk material\cite{Metal95,Getal97,Misko01}. 
Many experiments and theoretical\cite{SP99,P00}
developments on mesoscopic superconductivity have been made,
unravelling new physics related to the confinement of the
condensate. Experiments with single small particles\cite{Ralph97}, 
thin wires\cite{TinkhamQPS,TinkhamQPS2}, carbon
nanotubes\cite{nanotube1,nanotube2}, or DNA
molecules\cite{dna1,dna2,dna3} have been reported. Very clever 
solutions have been
given to the difficult problem of the contacts (see e.g.
\cite{dna2}), although it remains one of the main limitations in
the operation of these small systems\cite{contacts,nanotube3}.

The invention of the scanning tunnelling microscope\cite{Binnig},
STM, has been a breakthrough towards our control of the nanoworld.
Following this invention, several tools have been developed
extending the initial STM capabilities. Atomic force microscopy
\cite{BinnigAFM} has proven to be a powerful tool to investigate
both fundamental problems and others with extraordinary
technological importance, such as friction, wear or fracture.
Imaginative combinations of the working principles of both
techniques have promoted new tools for specific experiments.
Among them, we highlight the results by Rubio-Bollinger et
al.\cite{RubioChains}, that were able to measure force and
conductance simultaneously, extracting one atom after the other
from a surface, and creating the smallest and thinnest arrangement
of atoms ever made, an atomic chain.  Magnetic force microscopy
\cite{Martin} and scanning Hall probe microscopy\cite{SJB} are other 
useful members of this toolkit, whose
main achievements rest on the impressive control of the
displacements that can be done through the piezoelectric
deformation of some ceramic materials. This control is magnified
at low temperatures where atomic mobility is very low and the
creep effect in the piezoelectric ceramics is also reduced to a
very low level\cite{VieiraPiezo}.
There are many relevant
achievements in condensed matter physics that have appeared in the
twenty years span since the STM invention. One of these, related
to the main topic of this review, was the observation of an atomic
jump to contact when a metallic atomically sharp tip was carefully
approached to a sample of the same
material\cite{Gimzewski}. 
After this pioneering
experiment many others have been done to study transport and
mechanical properties of atomic size contacts using an
STM\cite{ReviewNic}. Taking advantage of the unprecedented
capability of control that STM has on the displacements,
nanometric indentations of the tip in the sample surface can be
made to create bridges of variable minimal
cross-section\cite{Agrait93,GarciaPRL,TSF94,GarciaScience}.

We will review here charge transport through superconducting
nanobridges and related structures, and the physical information contained 
on this transport.
Transport regime can be dramatically
modified by small changes in the minimal cross-section region, the
neck,  but the overall nanostructure (nanobridge)
remains unmodified when scanning through these regimes in the experiment.
At a high level of current, heating and other nonequilibrium
effects appear. In atomic size contacts  superconductivity and
quantum transport phenomena can be studied in a well-controlled
manner. Breaking the tip into two parts results into two atomic
size nanotips. One of these can be in-situ transported elsewhere
and used to make atomic resolution microscopy and spectroscopy
over a sample, without change in vacuum or temperature
conditions\cite{Rnbse04}. The application of an external magnetic
field confines the condensate around the bridge region creating a
nanoscopic superconductor with perfect interface with the normal
region, solving in a natural way the contacting problems
\cite{contacts} associated to this kind of structures. This unique
system gives us the possibility to make experiments in a highly
controlled situation. Theoretical calculations using
Ginzburg-Landau theory and Usadel equations provide a framework to
understand the most important aspects  of superconductivity in
these bridges.

We discuss first how the nanostructures are built and
characterized (section II). Then, in section III, we will review
theoretical models and experimental results about the transport
properties of these systems at zero magnetic field. We discuss
separately the three different conduction regimes: tunnel, atomic
contact, and low resistance ballistic transport. The same scheme
is used to discuss the transport properties in an applied field
(section IV). We conclude, in section V, with comments on open
questions and future studies which can be addressed with the
systems described here.

\section{Fabrication and characterization of the superconducting nanostructures}

\begin{figure}[p]
\begin{center}
\resizebox{0.7\columnwidth}{!}{%
\includegraphics{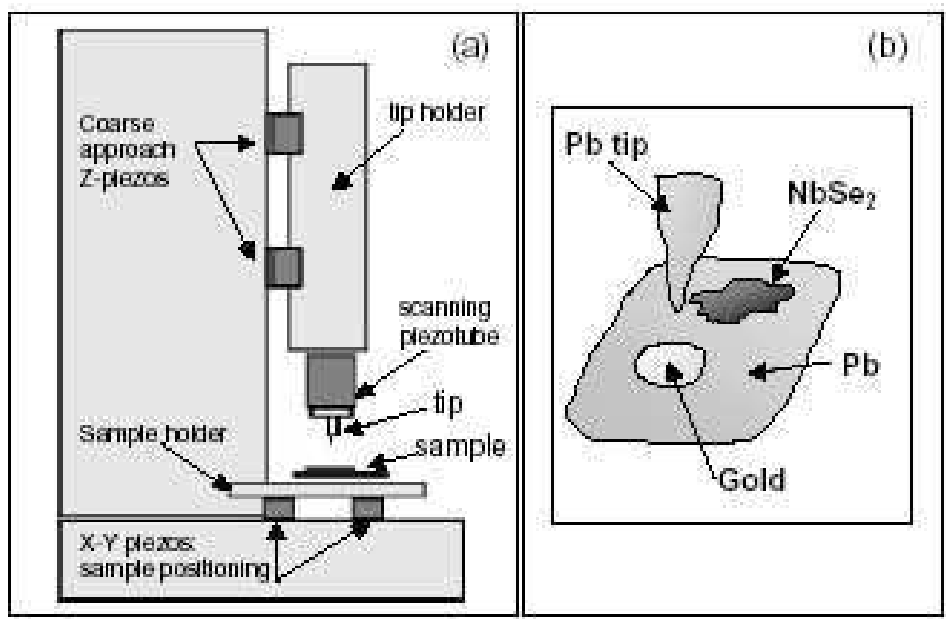}}
\caption {Scheme of the STM unit (a) and its composite sample
holder (b) used in the 
low temperature laboratory of the Universidad
Autonoma de Madrid.} 
\label{fig:stm} 
\end{center}
\end{figure}

The scanning tunneling (STM) \cite{Binnig} and the atomic force
 (AFM) \cite{BinnigAFM} microscopes, as well as some related techniques, are
versatile tools to penetrate in the nanoworld realm. The STM
allows to study the topography and electronic properties of a conducting surface
 with atomic spatial resolution. In the little more than
twenty years elapsed since its invention this technique has became
widely used. These instruments can be obtained from commercial 
suppliers, some
of them designed to work at low temperatures. However, home made
STM are in use in many laboratories, as they give the required
versatility and accuracy for doing specific research. Some home
made STMs are well adapted to be mounted in the cryogenic ambiance
of $^3$He-$^4$He dilution and $^3$He refrigerators, and to work
under magnetic fields\cite{Hess90,Stip1,Suderow00b,Moussy01,Suderow02b}. 
A cylindrically symmetric design is best suited for that. In
Fig.\ref{fig:stm}(a) we show a sketch of the STM built and used in
the low temperature laboratory of the Universidad Autonoma de
Madrid\cite{Martinez00}, and highlight its original aspects. The
coarse approach system, a piston whose controlled
movement is produced by piezoelectric stacks, is designed to make,
if wanted, strong indentation of the tip in the sample surface.
With this system, tip and sample can be approached from distances
of several millimeters in-situ at low temperatures. A piezotube
with capabilities of vertical displacements, at cryogenic
temperatures, in the range of several tenths of microns is used
for the fine control and movement. The table  on which the sample
holder is located is the other important part of
this instrument\cite{Martinez00}. At low temperatures it can be moved in the x-y
plane distances in the millimetric range using piezoelectric
stacks. This movement is well controlled and reproducible, and
allows to access with the tip a wide surface area within the same
cooling down run. Therefore, the sample holder can include a composite
sample of different materials, which can be studied together (see
Fig.\ref{fig:stm}(b)).

\begin{figure}[p]
\begin{center}
\resizebox{0.7\columnwidth}{!}{
\includegraphics{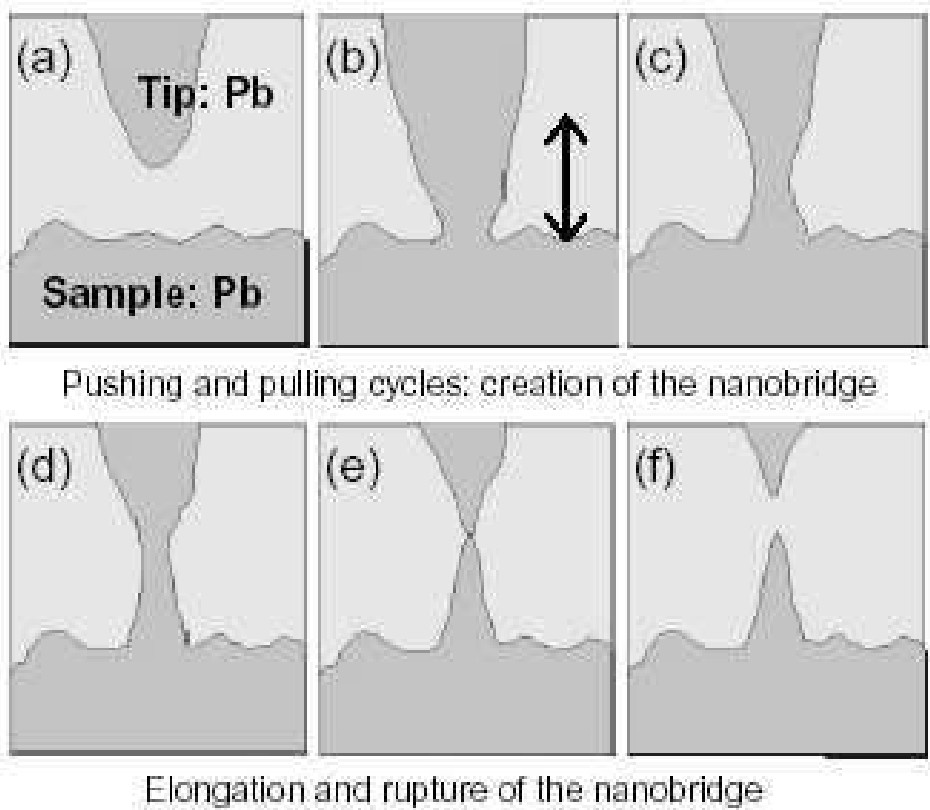}}
\caption {Sketch of the nanobridge fabrication process. Frames (a)
to (f) illustrate different stages of the process:
(a)Tip and sample in tunneling regime; (b) The tip
is pressed against the substrate, both electrodes deform
plastically and form a connective neck; 
(c) to (e) Indentation-retraction cycles produce a plastic 
elongation of the neck; (f) the rupture of the nanobridge
takes place.}
\label{fig:cartoon} \end{center}
\end{figure}

The structures that we discuss in this review have
been obtained using superconducting materials and the STM as a tool for
its fabrication. We call them nanobridges,
because the dimensions of the largest ones are a few hundreds of
nanometers. The fabrication of large
nanobridges works well with ductile metals like Au, Pb, Al and Sn
\cite {TSF94,Agrait95,UR97,ReviewNic} and semimetals, as it
is the case of Bi\cite{Rodrigo02}.
The first step of the
fabrication is to crash, in a controlled manner, a clean tip into
a clean substrate, normally both of the same material 
(see a schematic representation of the process 
in fig.\ref{fig:cartoon}). As the tip
is pressed against the substrate, both electrodes deform
plastically and then bind by cohesive forces,
 forming a connective neck (fig.\ref{fig:cartoon}(b)). Retraction of the tip results in the
formation of the neck that elongates plastically 
(fig.\ref{fig:cartoon},frames (c) to (e)) and eventually
breaks (frame (f)). 

Measuring the current, I, flowing through the neck at a fixed
bias, usually between 10 mV and 100 mV, as a function of the
displacement, z, of the tip relative to the substrate, it is
possible to follow the evolution of the neck.
These I-z curves are staircase-like
and strikingly reproducible when the process is
repeated many times\cite{Agrait93,UR97}.
The detailed analysis of the last steps from these experiments,
close to the breaking point of the nanobridge, is often
represented as conductance
histograms\cite{ReviewNic}. From those it has
been possible to extract, for some simple metals, relevant
information on quantum transport through atomic size contacts.

It was soon understood that the staircase shape of the I-z curves
reflected the sequence of elastic and plastic deformations
followed by the nanobridge \cite{Landman,Stafford}.
Only the minimal cross section, which determines the conductance 
and the current I at a fixed voltage, is modified.
This is a natural
result, as the stress is mostly concentrated around the narrowest 
part of the nanobridge, the neck. 
The conclusive evidence came from the combined STM-AFM
experiments\cite{TSF94,Agrait95,Rubio96}, in which the conductance and the forces
which develop during the elongation or contraction of the
nanobridge were simultaneously measured. The intimate relation between 
conductance steps and
atomic rearrangements was then established definitively.
These experiments were made with lead and gold, being the noble metal
the material most thoroughly studied. It was even possible to
observe how during the elongation of the bridge, gold deforms
plastically down to the last atom contact, and chains consisting of
several atoms were created\cite{Yanson98,RubioChains}.

The
conductance observed for these gold atomic contacts is quite close to
1G$_0$, where $G_0 = 2e^2 /h$  is the value of the quantum of
conductance\cite{Landauer,W88}. The force involved in the rupture 
of these one-atom
contacts is also well defined, with a value of 1.5$\pm$0.1
nN\cite{Rubio96,RubioChains}. Transport experiments in several
other elements in the superconducting state
(Pb, Al or Nb, and also in Au, made superconducting using the
proximity effect), have permitted to establish a clear
relationship between the conductance of the last contact and the
chemical nature of the atom
involved\cite{Setal97,Setal98,Rubio03}.
 Along with these experimental achievements there have
been important and successful efforts to get a theoretical
understanding of this subject. A recent review by 
Agra\"{\i}t et al.\cite{ReviewNic} provides
a comprehensive vision of this field.

\begin{figure}[p]
\begin{center}
\resizebox{0.7\columnwidth}{!}{%
\includegraphics{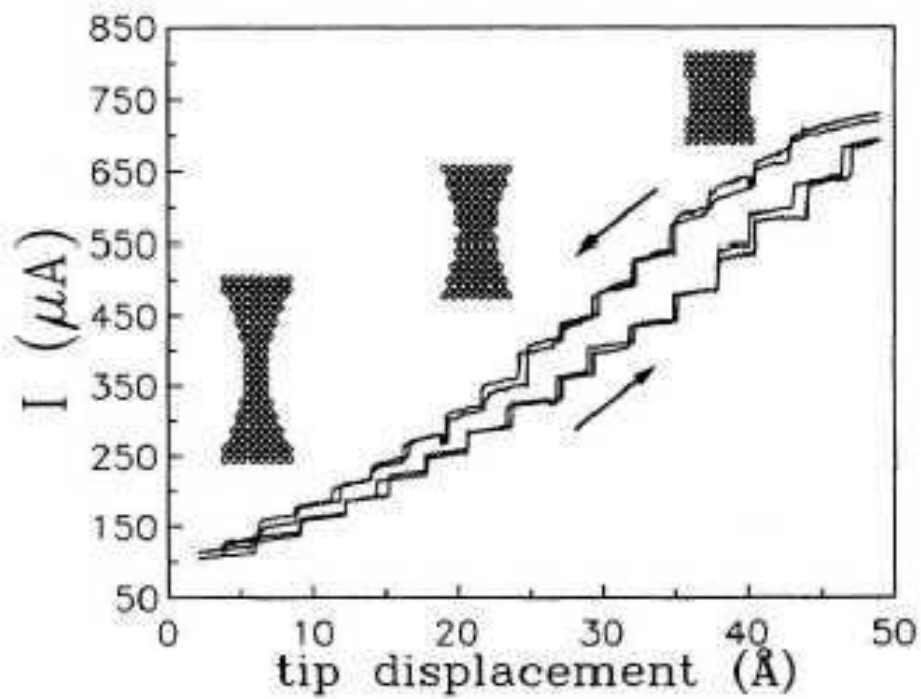}}
\caption {Experiments made at 4.2 K, using Pb tip and sample, from
Ref.\protect\cite{Agrait93}. The current for a bias of 50mV was
measured changing the area of the contact. A reproducible and
regular structure in the current vs tip displacement curves
develops. Remarkably, this high reproducibility is obtained in a
process involving plastic deformations.
The estimated shapes at different stages of the nanobridge
along the I-z curves are sketched.} \label{fig:cuello93}
\end{center}
\end{figure}

\begin{figure}[p]
\begin{center}
\resizebox{0.7\columnwidth}{!}{%
\includegraphics{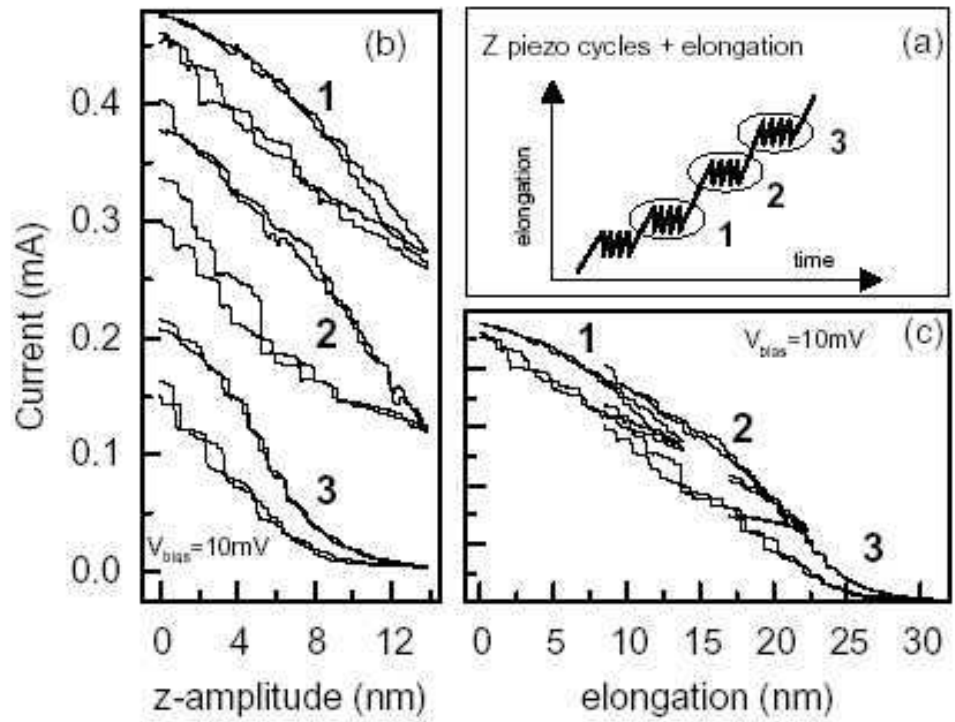}}
\caption {(a) Schematic indication of the time evolution of
the process of elongation
of the nanobridge.
(b) I-z curves as recorded in a real nanobridge fabrication. 
Each group of curves (1,2 and 3) correspond to a different stage
of the elongation of the nanobridge, i.e., a different neck.
These curves can be arranged in order to account for the
total elongation of the nanobridge (c).
(Data from \protect\cite{Rodrigounpub}) }
\label{fig:serieIZ} \end{center}\end{figure}

Here we are mainly concerned with the overall shape of the
nanobridges created with the STM and, in particular, using superconducting
materials. The results shown in fig.\ref{fig:cuello93} 
were the first clear
indication that the controlled fabrication of superconducting nanobridges
using a STM as a tool and as a probe was possible\cite{Agrait93}.
In this experiment,
made at 4.2K, high purity lead (T$_c=$7.14 K) was used for tip and
sample. The current for a bias of 50mV was measured changing the
area of the contact. Both quantities, minimal cross section and current can
be related using the simple Sharvin formula\cite{Sharvin}:
\begin{equation}
G_S=\frac{2e^2}{h}\frac{k_F a}{2}^2 
\label{Sharvin}
\end{equation}
\noindent where $G_S = dI/dV$ is the conductance, $a$ is the
radius of the contact, $k_F$ is the Fermi wave vector, $h$ is Planck's constant 
and $e$ is the electron charge. This expression is strictly valid for ballistic
transport (i.e., electronic mean free path $\ell \gg
a$) \cite{ReviewNic}.
 Assuming that deformations are confined to
a small region of volume around the narrowest cross section, and
that the neck is parabolic , the evolution of its shape can
be obtained from the measured I-z curves (Fig.\ref{fig:cuello93}).
Large nanobridges could be obtained with the procedure
schematically represented in fig.\ref{fig:serieIZ}(a).
Following a strong
indentation the tip is receded while moving back and forth with a
smaller amplitude without breaking the contact. Then a
reproducible and regular structure in the current vs tip
displacement curves develops.

Untiedt et al. \cite{UR97} developed a slab model suggested by
the results of combined STM-AFM experiments. When the
conductance is rather constant, the force varies linearly, while
the abrupt jumps in conductance are correlated to abrupt force
relaxations. Between the relaxations deformation is elastic so
that no energy is dissipated. The nanobridge is modelled as a
constriction with cylindrical geometry, consisting of slabs of
different radii and thickness, symmetrical with respect to 
its minimal cross section. The elastic properties of the
nanobridge, e.g., the Young's modulus $E$ and Poisson's ratio $\mu $ are
considered identical to the bulk values. The basic assumption for this model is
that only the narrowest part of the nanobridge, the neck, deforms plastically. 
This assumption could break down for
temperatures larger than about 50\% of the melting temperature,
for which diffusion will be important\cite{UR97,Gay02}, but it is valid at the
temperatures of interest for this review, where atomic mobility is
negligible. The slab model provides a good description of the shape and dimensions of the
scanned feature, left onto the surface of the sample after breaking a fabricated
nanobridge.
The atomic sharpness of the tips obtained using this method, permits to obtain images with atomic
resolution \cite{Rnbse04}. A composite sample like
the one shown in fig.\ref{fig:stm}(b) was used. After preparing an atomically sharp Pb tip
on the lead surface, the sample holder is moved so that the
NbSe$_2$ single crystal surface could be reached and scanned by the tip.

Large amplitude phonon peaks were observed in point contact spectroscopy
experiments in long nanobridges with I-z curves showing prominent and
repetitive stepped structure\cite{UR97}. This observation was interpreted as
an indication of crystallinity due to ``mechanical annealing'' of the defects
by repeated plastic deformation.

\newpage

\section{Transport regimes in superconducting nanostructures at zero magnetic field}

\subsection{Theory}

The current flowing through a tunnel junction (see ref.\cite{Wolf}
 for a general review on tunneling spectroscopy) is
given by the convolution of the density of states (DOS) of both
electrodes:

\begin{equation}
I=\frac{G_N}{e} \int_{-\infty}^{\infty} N_{1}(E) N_{2}(E)
 \left [ f(E) - f(E+eV) \right
] dE
\end{equation}

\noindent where $N_{1}$ and $N_{2}$ are the normalized density of
states and $G_N$ is the normal-state conductance of the junction.

For BCS superconducting electrodes, the density of
states takes the form\cite{BCS57},
 $N(E)=Re\left[ E/\sqrt{E^2-\Delta ^2}\right] $,
being $\Delta $ the superconducting gap.
If tunnelling is
performed between two identical superconductors, at zero
temperature no current can flow for voltages $V< 2\Delta /e$. At
finite temperatures, due to thermal excitations, states above the
Fermi level can be populated and those below depopulated, allowing
finite quasiparticle current flow at voltages smaller than $2\Delta /e$.

In tunnel junctions where the barrier is sufficiently low,
multiple scattering of Cooper pairs leads to a finite conductance
below the superconducting gap . In these processes,
generically called Andreev reflection\cite{A64}, an electron is
reflected as a hole at the junction, leading to the transmission
of a Cooper pair\cite{AVZ78,AVZ79a,AVZ79b,Z80,G89}. Subgap Andreev
reflection takes place both in normal-superconductor junctions and
in superconductor-superconductor junctions. A quantitative
analysis of these processes in normal-superconductor (N-S)
junctions,
was made in ref.\cite{BTK82}, and later extended to
superconductor-superconductor junctions\cite{Oetal83}, where the
junction was described as two superconductor-normal junctions in
series.
A complete determination of the transport properties of a
superconductor-superconductor junction at finite voltages requires
to take into account not only the d.c. current, but also the
higher harmonics.
A detailed analysis of the time dependent
current flowing under an arbitrary applied voltage, taking into
account all multiple Andreev reflections (MAR), for a single
channel through the junction, was done in ref.\cite{AB95} and in
ref.\cite{CML96}.  At large voltages $V\gg 2\Delta/e$, I-V curves are linear, with a slope given
by the normal state conductance, but do not extrapolate to zero. The excess
current\cite{Octavio78} is defined by
\begin{equation}
I_{exc}=lim_{V\rightarrow \infty}[I(V)-I_n(V)]
\label{exceso}
\end{equation}
with $I_n$ the current in the normal state. At low voltages $V\leq 2\Delta/e$,
the I-V curves are strongly non-linear, showing inflections at $eV=2\Delta/n$. These
features, known as Subharmonic Gap Structure (SGS), are a consequence of
the multiple Andreev reflections.
They can be modified
by the internal structure of the junction, or when the
transmission coefficient has a significant energy
dependence\cite{BG02}.

The non-linearity of the $I-V$ curves has been used to discuss the
contribution from different channels in junctions of atomic
dimensions\cite{Setal97,Setal98}.
The current in a
superconducting-superconducting constriction can be written as the
sum of the contribution of $N$ channels in parallel
$I=\sum_{n=1}^{N}i(V,T_n)$\cite{Bardas97}.
The current carried by each channel is
the corresponding to a one dimensional superconducting
constriction with transmission $T_n$. In the normal state, the
total current depends only on the total conductance, independently
of the transmission of the individual channels which contribute to
it.
This is not the case in the superconducting state.
In an m-th Andreev reflection process the barrier is
transversed $m+1$ times. The probability that this process occurs
scales as $T_n^{m+1}$. 
Hence, the total current $I$ strongly dependens on the
set of individual transmission coefficients.

The current through a superconductor-superconductor (S-S) junction
shows other features due to the phase rigidity of the condensate.
The most striking manifestation of this property is the Josephson
effect\cite{J62,J65,Barone}. A current below a 
certain value, $I_c$, can
flow between two superconductors at zero voltage.
Following Ambegaokar and Baratoff\cite{AB63},
the critical current, $I_c$,  of a tunnel junction between BCS 
superconductors can be written as 
$I_c \approx ( \pi G_N \Delta) / ( 2 e )$ . 
This analysis was later
extended to other types of junctions\cite{L79}.
The value of the critical current $I_c$ for a short and narrow constriction
was calculated by Kulik and Omel'yanchuk\cite{KO1} in the case of a point
contact much wider than the Fermi wavelength, when the quantization of the
momentum can be neglected. Its value for a quantum point contact with a small
number of conducting channels was calculated by Beenakker and van
Houten\cite{Beenakker91c,Beenakker91b}. 
The observation of Josephson current is affected by the balance between 
the thermal energy, $k_BT$, and the
Josephson coupling energy\cite{AB63}, given by
\begin{equation}
E_J=\frac{\Delta R_Q}{2 R_N} \label{Josephson}
\end{equation}
\noindent where $R_N$ is the normal state resistance and 
$R_Q=h/4e^2 = 6.45k\Omega $.
For resistances such that the Josephson coupling energy is 
comparable to the thermal energy, 
the superconducting phase dynamics is dominated by thermal fluctuations, 
making the Josephson current to appear as a peak centered at small finite voltage.
In this case the phase motion can be viewed as diffusive. The
I-V characteristics of such a junction have been calculated
 by several authors \cite{Ivanchenko,Harada,Kautz,Ingold,Vion}
 using the washboard potential model\cite{T96}.

\subsection{Experiment}

The method described in section 2 has been used 
(see ref.\cite{REPJ04} and references therein) to create 
superconducting tips made
of lead and aluminum, with transition
temperatures of 7.2K and 1.2K respectively.
We will review now the different transport regimes
(tunneling, atomic contact and weak-link) 
accesible through the fabrication and rupture of a superconducting nanobridge. 
Along this process it is
possible to follow in detail the evolution of the I-V
characteristics in a wide range of conductance.
Fig.\ref{serieG} shows the typical
evolution of the conductance spectra, $dI/dV$ vs $V$, as the 
junction resistance 
 is varied from vacuum tunnelling regime, $10
M\Omega $, to a point contact regime with $100 \Omega $.
These STS measurements were done at 1.8 K using tip and sample made of lead.

\subsubsection{Tunnelling regime}

Many spectroscopic experiments made with STM on superconductors
have shown I-V curves with notable differences with respect to the
expected behavior for a BCS superconductor. It was suggested that
the high density of current through the atomic size constrictions
could break Cooper pairs, inducing a smearing in the spectroscopic
curves\cite{Renner91}. However, it appears that this does not
influence the sharpness of the obtained spectra, as emphasized in
\cite{Suderow01,Moussy01,Suderow04,REPJ04}. Criteria to
test the effective resolution of the STM experimental setup have been 
discussed in refs. \cite{Suderow04,REPJ04}, based on
measurements on aluminum in $^3$He/$^4$He dilution refrigerators.

Aluminum is considered
in many aspects the archetypical weak coupling BCS superconductor, with a well
defined value of the superconducting gap.
Ideal I-V curves in the S-S tunnelling
regime at very low temperature present the well known features of
zero current up to the gap edge at 2$\Delta $, where there is a
jump to non-zero current\cite{Wolf,Mersevey94}. This appears in
the tunneling conductance curves ($dI/dV$ vs $V$) as a divergence
at energy 2$\Delta $, present at all temperatures, which is the
sharpest feature that can be observed in tunnelling spectroscopy
measurements in superconductors.
Therefore, the measurement of the current in the tunnelling regime
is a direct test of the energy
resolution of the experimental set-up. This energy resolution can
be introduced in the calculus of the curves as a narrow gaussian
distribution, which simulates the noise in the voltage source, and
has a halfwidth in energy of $\sigma $ \cite{REPJ04}.
 
\begin{figure}[p]
\begin{center}
\includegraphics[width=0.7\linewidth]{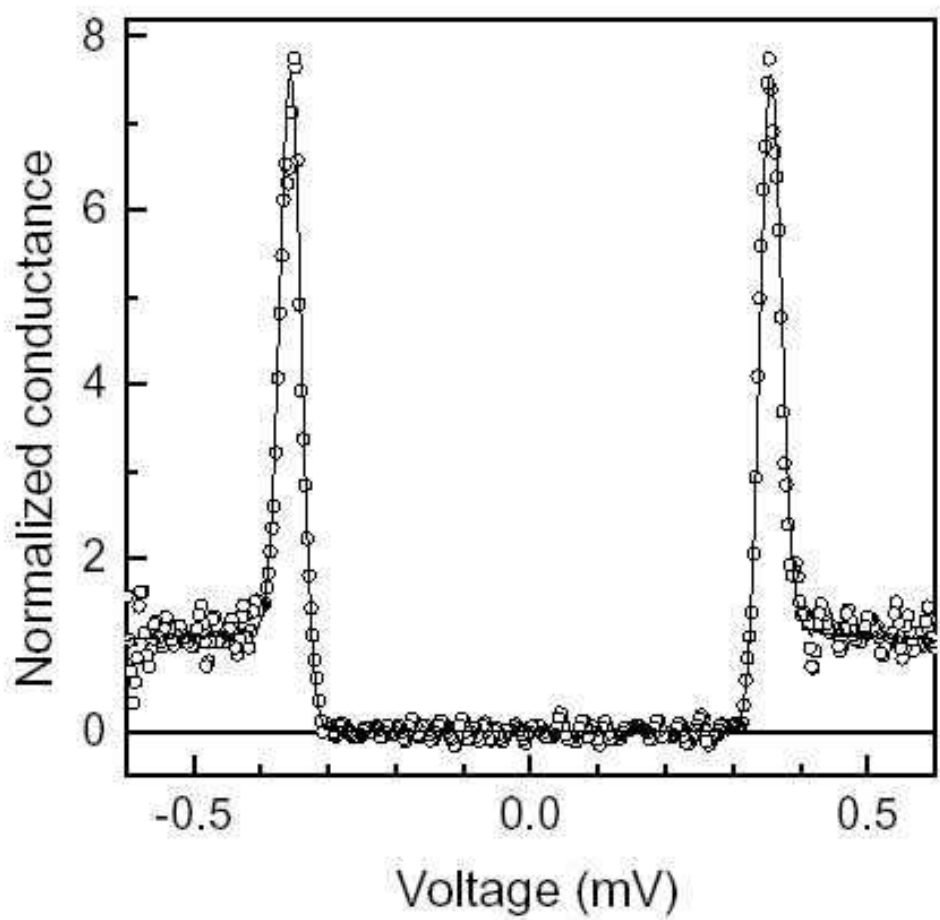}
\caption{ Al-Al tunnelling
conductance curve obtained at 70 mK and its corresponding fitting.
Data taken from
\protect\cite{REPJ04}.} \label{Altunel}
\end{center}
\end{figure}

Dynes et al.\cite{Dynes78} introduced a
phenomenological broadening parameter, $\Gamma $, into the BCS density of states
to account for the broadening of the gap edges in the spectra of 
dilute bismuth alloys in lead, as
due to finite lifetime effects of the quasiparticles. 
This lifetime broadening model has
been applied routinely to situations in which the main source
of smearing or broadening of the spectra is of experimental
origin.
Fig.\ref{Altunel} presents an experimental Al-Al tunneling
conductance curve measured in a dilution refrigerator, and the
corresponding fitting\cite{REPJ04}. The calculated curve was
obtained with the parameters $\Delta =175\mu eV$, $T=70mK$ (base
temperature of the system) and energy halfwidth $\sigma =15\mu
eV$. At non-zero temperature the current expected for
these junctions at subgap energies is not zero, due to the thermal
broadening of the Fermi edge. However, at low temperatures, this
current disappears exponentially and it is hardly detectable.
Within an experimental resolution in current of 1 pA,
the same curve is obtained up to $250mK$.

Tunneling experiments using superconducting tips obtained from 
lead nanobridges fabricated with the STM 
have been reported\cite{Rnbse04,REPJ04}.
Lead is a
strong coupling superconductor, and it was found since early
tunnelling experiments\cite{Pb70s1,Pb70s2,Pb70s3} that its gap
value is not constant over the Fermi surface. Recent results give new
support to this scenario\cite{Short00}.
The tunnelling curves

obtained at 0.3K (fig.\ref{Pbtunel1}(a)) 
present coherence peaks with a finite width, 
shown in detail in fig.\ref{Pbtunel1}(b), 
larger than the one expected considering only the resolution in
energy. This additional width is a consequence of the gaps distribution in lead.
The conductance curves
are flat bottomed at a
zero value of the conductance inside the gap region, indicating the absence
of a relevant finite lifetime source. 
Therefore, the experimental spectra were simulated by means of a 
gaussian distribution of gap values, as well
as a similar distribution accounting for the energy resolution of
the spectroscopic system. To fit the experimental data
for lead, the temperature (0.3K) and energy resolution ($\sigma
=20\mu eV$) were kept fixed, leaving the superconducting gap,
$\Delta $, and the halfwidth of the distribution of values of the
superconducting gap, $\delta $, as free parameters, obtaining
$\Delta =1.35meV$ and $\delta =25\mu eV$. A detailed discussion of
this analysis can be found in Refs.\cite{Rnbse04} and
\cite{REPJ04}.

As lead is a strong-coupling superconductor, the features due to
phonon modes are observed in the tunnelling conductance
curves (fig.\ref{Pbtunel1}(a)). According to the well-known
properties of strong-coupling superconductors \cite{Eliashberg,MCMR}, a
peak in the effective phonon spectrum gives a peak in the voltage
derivative of the conductance located 
at $\Delta_0+\omega _{L,T}$, being $\omega _{L,T}$ the energies 
corresponding to the phonon modes. 
No significant difference neither in the value
of the superconducting gap nor in the phonon modes ($\omega _{T} =
4.4 meV$ and $\omega _{L} = 8.6 meV$) with respect to planar
junction experiments is found within the experimental resolution.
The progressive fading out of phonon features in the S-S tunnelling
conductance curves, as both electrodes are approached towards
contact was discussed by Rodrigo et al. \cite{JGRfon94}, being a
consequence of the mixing of spectroscopic information from
different energies close to Fermi level as multiple Andreev
reflection processes become more important to the total
conduction.

Quasiparticle tunnelling is not the only contribution to the total
current. It is also possible to observe
tunnelling of Cooper pairs, the Josephson effect. As noted in
refs.\cite {Stip1,Stip2,Stip3,Stip4,Balatsky}, the measurement of
the Josephson effect in atomic size and high resistance vacuum
junctions is a
true challenge. In a typical tunneling experiment,
with normal-state resistances in the M$\Omega $ range, and not
very low temperatures, the thermal energy $k_BT$ is higher
than the Josephson coupling energy $E_J$ (eq. \ref{Josephson}). 
For Pb junctions with a normal-state resistance of
1M$\Omega$, both energies are similar at 50 mK. For thermal
energies bigger than, but comparable, to the Josephson binding
energy, pair tunnelling would be observed, but the pair current
will be dissipative, i.e. with the voltage drop proportional to
the rate of thermally induced phase slips across the junction \cite{Stip2,Stip3}. 
Experiments on ultrasmall Josephson junctions have shown that 
the Ambegaokar-Baratoff critical current
can be reached at low temperature, if the junction
is placed in an appropriate controlled 
electromagnetic environment\cite{Joyez01}. 

By reducing the distance between tip and sample, it is possible to
cover a wide range of resistance and temperatures\cite
{Stip2,Stip3,JGRJJ94,REPJ04}, and  to get information
on different Josephson regimes by changing in a controlled way
the ratio between thermal and Josephson binding energies.
The increase of the
Josephson current as the tunnel resistance is decreased is shown in
the inset of Fig.\ref{ingn}(a). This effect appears as an
increasing peak at zero bias in the conductance curves, observed
in the lower curves of Fig.\ref{serieG}, which are normalized and
blown up in Fig.\ref{ingn}(b).

It is important to remark that only the precise determination of
the limits in spectroscopic resolution permits to extract relevant
information, as the one about the gap distribution in lead, from
local tunneling experiments. Recently there have been several
reports on new superconducting materials which indicate that a
single gap in the Fermi surface is not the more frequent case\cite
{Rubio01,Taillefer03,Yokoya03,REPJ04,Suderow04}. Multiband
superconductivity and gap anisotropy seem to be more ubiquitous
than previously thought. These observations enhance the importance
of precise local tunnel measurements to shed light on a variety of
open problems.

The STM superconducting tip resulting from the rupture of a
nanobridge, in situ at low temperatures, 
has been used recently\cite{Rnbse04,REPJ04}
to obtain spectroscopic information and topographic images with
atomic resolution on other samples. This was possible by using
sample holders like the one described in section 2. Other
simultaneous STM/STS experiments using superconducting tips were
reported in the past. 
A Nb tip, previously cleaned at low temperature by field emission, 
was used by the authors of ref.\cite{Stip1} to
perform STS on a NbSe$_2$ sample, whose surface was imaged with
atomic resolution. A different approach is described in ref.\cite{Stip5}, 
where a controlled Pb/Ag proximity bilayer was deposited onto precut Pt/Ir
tips to obtain STM superconducting tips suitable for STM/STS experiments.
Finally, already in 1994,
a surface of lead was scanned at 4.2K using a tip
of the same element resulting from the rupture of a nanobridge, and
spectroscopic measurements at different conductance regimes were
performed\cite{Rodrigo94}.

\subsubsection{Atomic size contact regime}

\begin{figure}[p]
\begin{center}
\includegraphics[width=0.7\linewidth]{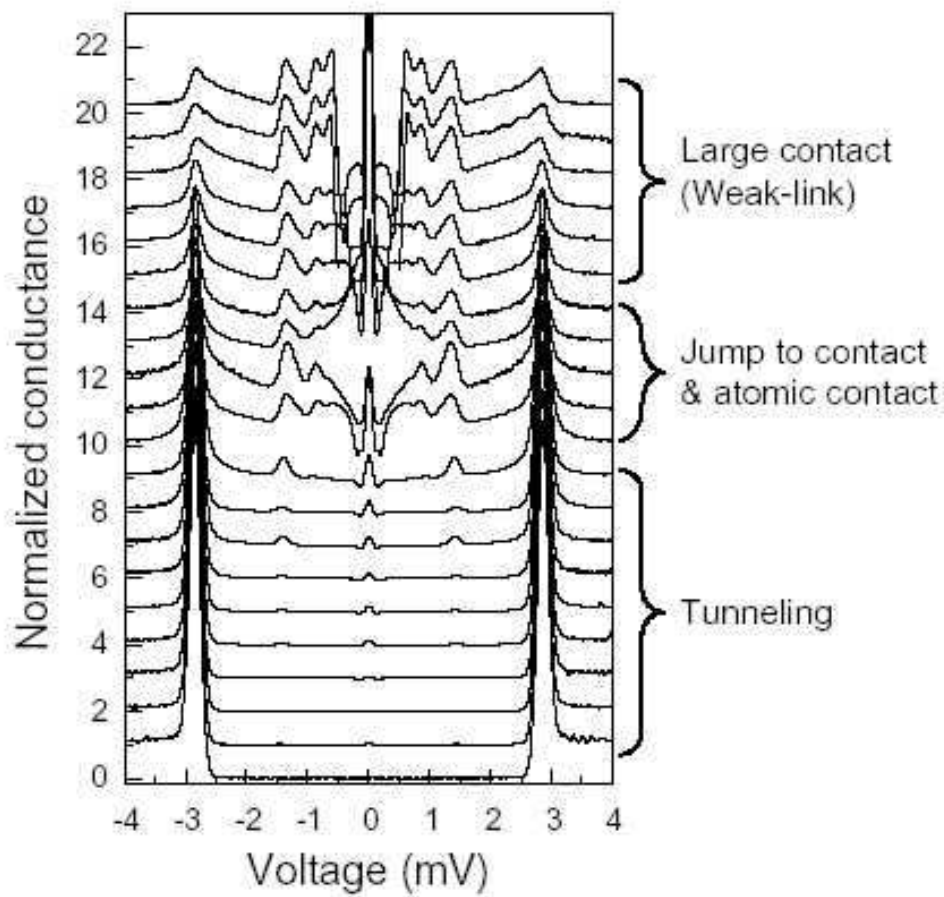}
%\resizebox{0.7\columnwidth}{!}{%
%\includegraphics{serieG.eps}}
\caption {Evolution of the conduction spectra along the different
stages of the creation of a Pb nanostructure (nanobridge). For a given
nanostructure, it covers resistances from $100 \Omega $
(large point contact) up to $10 M\Omega $ (vacuum
tunneling). Atomic contact takes place for resistances about $10 k\Omega $.
 Measurement performed at 1.8K. Data from
Ref.
\protect\cite{Rodrigounpub}
.} 
\label{serieG}
\end{center}
\end{figure}

As the two parts of the nanobridge are approached, 
the transmission probability
through the barrier increases, and MAR lead to SGS at
voltages $V \leq 2\Delta /e$, and to an excess current at large
voltages. 
The appearance of SGS can be seen in fig. \ref{ingn}, both in the
current in the inset in (a) and in the normalized conductance in (b).
These effects are observed in the curves in figs.\ref{serieG} and
 \ref{ingn}.
Fig.\ref{ingn}(a) presents the normalized $I-V$ curves ($I \times
R_N $ vs $V$) corresponding to all the measured range of
resistances showing the transition towards contact between tip
and sample, and the development of SGS features and the excess
current. The SGS features at $V =2\Delta /ne$ ($n$ = 1, 2, 3 and
4) which start to develop in tunneling regime, are finally clearly
seen in contact regime (see fig.\ref{serieG}). Peaks with high $n$ are
enhanced at higher conductance (higher transparency of the
barrier) as the probability of multiple Andreev reflections of
high order increases.

At present,  single atom
contacts can be achieved as a routine procedure.
The non linear I-V curves of these contacts
can be fitted to a sum of contributions from the different quantum
channels\cite{Setal97,Setal98}.
The number of conducting channels and their transmission are taken as fitting
parameters. The number of channels per atom depends on 
the chemical element.
The individual transmission set changes for different
contacts, as shown in fig. \ref{contacto1}(a).
Only three channels are needed to
fit these kind of curves corresponding to Pb contacts
\cite{Setal98,Suderow00c,Rnbse04}. This is taken as an indication that there
is a single atom contact between the nanostructures.

Josephson current has been measured 
in aluminum atomic point contacts containing a small number
of well characterized conduction channels\cite{Joyez00}. 
These contacts were made using microfabricated break junctions.
The authors found that the value of the supercurrent
is related to the dissipative branch of the IV characteristics,
like in usual macroscopic Josephson junctions,
although in the latter the contribution of the different channels
cannot be disentangled. This fact
strongly supports the idea of the supercurrent being carried
by Andreev bound states and show that the concepts of
mesoscopic superconductivity can be applied down to the
level of single atom contacts.

\subsubsection{Weak link regime}

\begin{figure}[p]
\begin{center}
\includegraphics[width=0.7\linewidth]{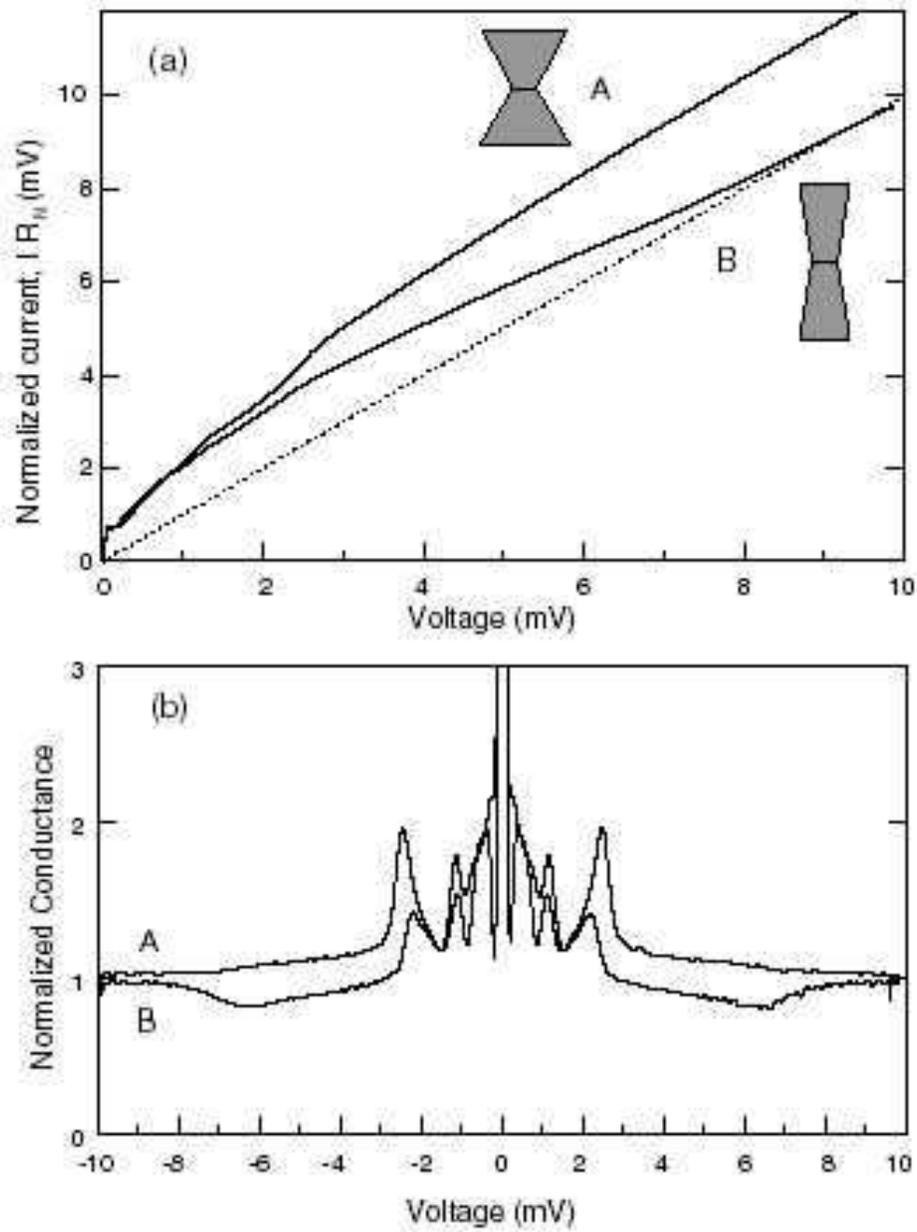}
\caption{ Normalized $I-V$ (a), and conductance (b) curves correspondng to two different
nanobridges with the same normal resistance, $500 \Omega$. Clear
heating effects appear for the long and
narrow bridge (B). The geometries of both
nanobridges are sketched.
The dotted curve in (a) indicates the normal state situation.
(Data from Ref.\protect\cite{Rodrigounpub}, see also
\protect\cite{Suderow00b}) } \label{JJheat}
\end{center}
\end{figure}

When a large point contact is formed between the two parts of the nanobridge,
 an interesting
phenomenology is found in the spectroscopic curves at zero field.
As the voltage increased, the system jumps out of the Josephson
branch and shows the previously discussed multiple Andreev
reflections.
Due to the strong current density the temperature raises locally around the
contact. The gap decreases and at high voltages the excess current is
lost\cite{Flensberg,JGRJJ94}. It vanishes completely when the system
reaches locally the critical temperature. This phenomenon appears
in the conductance curves as a bump after the SGS, and a recovery
of its normal state value when the critical temperature is
reached. Heating effects can be controlled in-situ
by changing the form of the neck. Long and narrow bridges will
show large overheating effects, whereas short and wide bridges are
easily thermalised and much larger voltages have to be applied to
observe overheating. Local overheating is also
commonly observed in classical point contacts
\cite{JansenPhon1,JansenPhon2,JansenPhon3}, although in that case
it is not possible to control its magnitude. Fig. \ref{JJheat}
shows a typical example of
conductance curves corresponding to two different nanobridges with
the same normal resistance, $500 \Omega$, equivalent to a minimal
cross section of about 20 atoms.
The geometry of the bridges can be checked through
the evolution of the $I-Z$ curves.
Clear heating effects appear in
the curve obtained for the long and narrow bridge (curves B) as
compared to the situation of a wide bridge (curves A): the SGS
peaks move towards lower voltage due to the decrease of the gap
value and there is a decrease of
conductance, indicative of the loss of the excess current.
Finally the conductance recovers its normal state value as the
local temperature rises above $T_c$. These effects are not present in
curve A.

\begin{figure}[p]

\begin{center}
\includegraphics[width=0.7\linewidth]{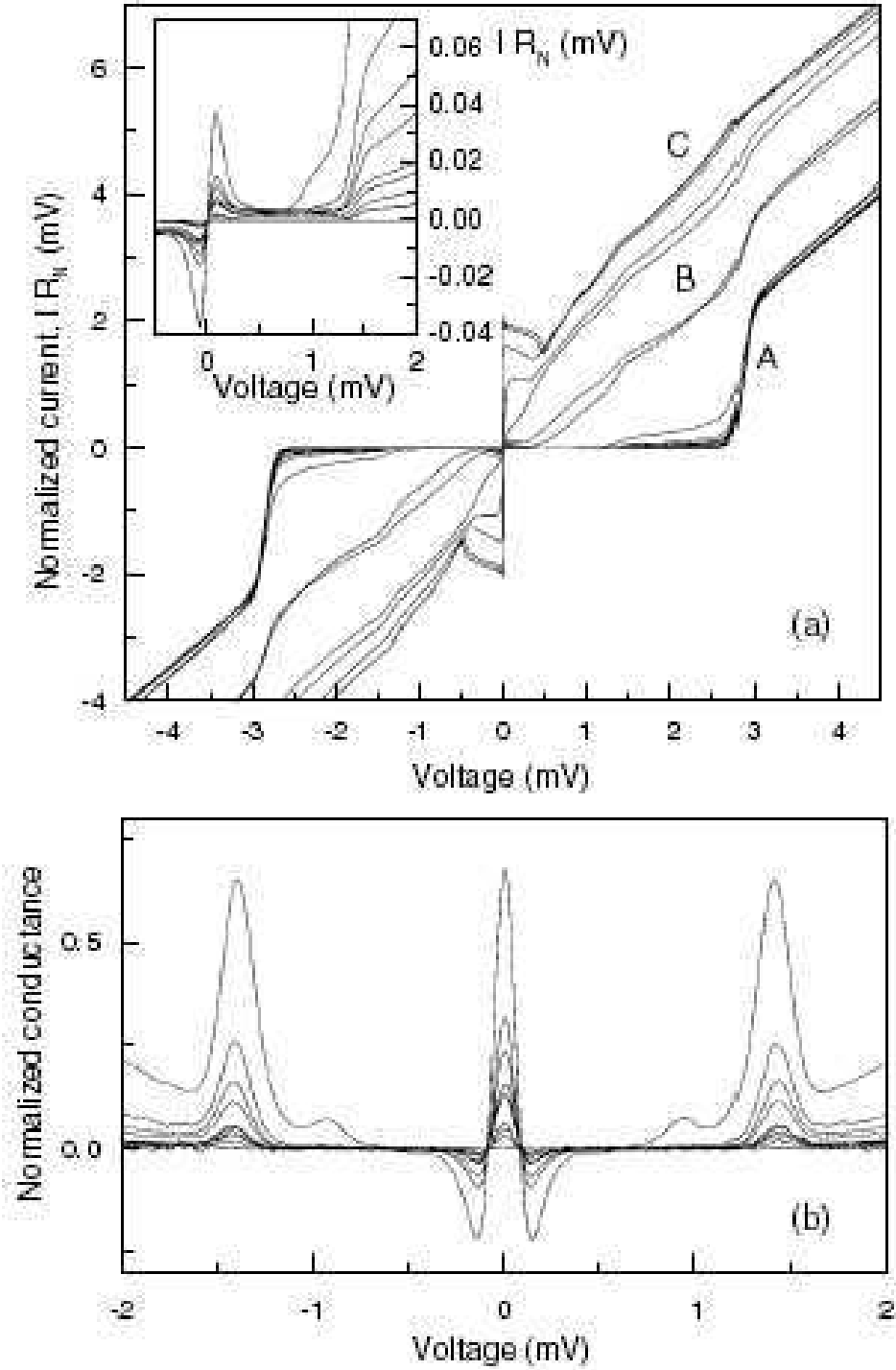}
\caption{ (a) Normalized $I-V$ curves ($I R_N $ vs $V$)
corresponding to all the measured range of resistances. The
different transport regimes can be identified: tunnelling (A), the
transition towards atomic contact between tip and sample (B), and
large contact regime (C). The SGS features and the
excess current develop along the transition from A to C. Inset:
blow up of the region close to zero bias for the curves in
tunneling regime.
 (b) Normalized conductance curves corresponding to tunnelling regime.
The inset and frame (b) illustrate how the balance between Josephson binding
energy and thermal fluctuations affect the observation of Josephson current in
tunneling regime. (Data from ref.\protect\cite{Rodrigounpub}, see
also ref.\protect\cite{REPJ04})} 
\label{ingn}
\end{center}
\end{figure}

Initial works on lead nanobridges fabricated with a
STM \cite{JGRJJ94} reported on the above mentioned heating effects,
as well as on the crossover between the 
Ambegaokar-Baratoff (tunneling)
and Kulik-Omel'yanchuk (weak-link)
limits for the transport of Cooper pairs.
Fig.\ref{ingn}(a) illustrates the difference in the 
normalized critical current (i.e. critical voltage) 
between the tunneling regime (curves A and inset) and 
the large point contact or weak-link regime (curves C).

\begin{figure}[p]
\begin{center}
\includegraphics[width=0.7\linewidth]{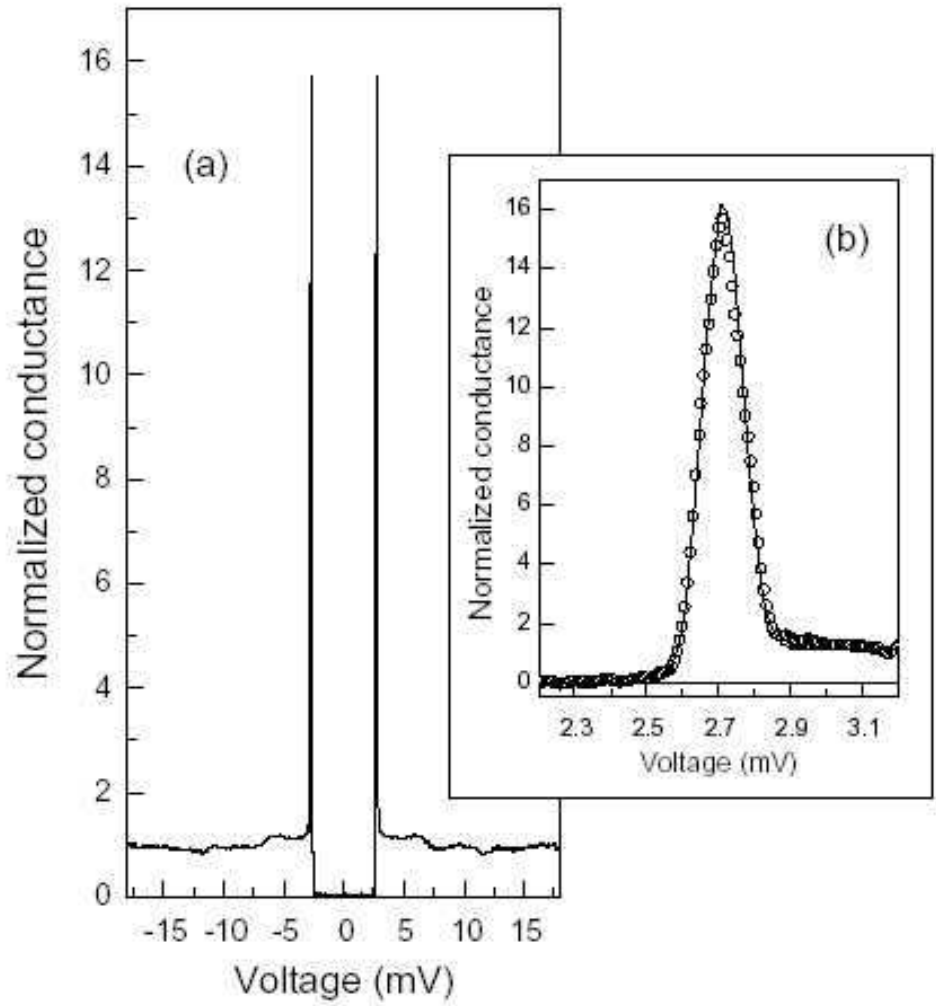}
\caption{(a) Tunnelling conductance curve obtained in a $^3$He
cryostat at 0.3K with tip and sample of Pb ($R_N=1 M \Omega $).
(b)Zoom of the gap edge.
The theoretical conductance (line) has been
calculated with the parameters described in the text, in order to
reproduce the experimental curve (circles). Data taken from
\protect\cite{REPJ04}.} \label{Pbtunel1}
\end{center}
\end{figure}

\newpage

\newpage

\section{Superconducting bridges under magnetic fields}

\subsection{Theoretical models}

As mentioned in the introduction, samples of dimensions smaller than or
of the order of the London penetration depth are superconducting
at magnetic fields well above the bulk critical field. The reduced
dimensionality blocks the creation of Meissner screening currents,
and the kinetic energy associated with them does not contribute to
the total free energy of the superconducting state. The lateral
dimensions of typical superconducting nanobridges are easily of
the order of, or smaller than, the London penetration depth
$\lambda$. In the case of Pb, which is the material most
intensively studied, the zero temperature limit of this quantity
$\lambda_0=32nm$. Therefore, the superconducting properties of the
nanobridges strongly depend on its geometry and on the magnetic
field. Several theoretical approaches have been used to describe
the experiments. All of them are valid for superconductors in the
dirty limit, i.e. with a mean free path $\ell$ smaller than the
superconducting coherence length $\xi$ (in Pb, $\xi_0=52nm$). This
assumption is easily justified by the typical lateral dimensions
of the nanobridge, which can be taken as a good measure of the
order of magnitude of the relevant mean free path
$\ell$\cite{UR97,Petal98}. The reduction of $\ell$ does not affect
the superconducting properties of isotropic, s-wave
superconductors\cite{A59}.

In uniform two dimensional or one dimensional structures, as thin
films or wires, the effect of a parallel magnetic field is well
described by the pair-breaking theory, reviewed in Ref.\cite{M66},
and originally developed to account for the effect of magnetic
impurities \cite{ABRIKOSOV}. Contrary to the simple BCS case, in
the presence of a pair-breaking mechanism the order parameter and
the gap in the spectrum are not equal, and gapless
superconductivity is found close to the critical field. As shown
below, a more elaborate treatment, the variable radius pair
breaking model (VRPB), is needed to account for the particular
geometry of real nanobridges, which are three dimensional
cone-like objects\cite{Suderow00c,B00}. The VRPB model uses
Usadel's formalism and gives the temperature and magnetic field
dependence of the density of states. It reduces to the
pair-breaking description\cite{M66} when considering a uniform
wire. Alternatively, the Ginzburg-Landau (GL) approach has been
used to obtain  information about the geometrical distribution of
the superconducting condensate in the nanobridge, its eventual
vorticity and the dependence of the critical current as a function
of field.

\subsubsection{Usadel approach. VRPB model}

The electronic structure of a superconductor can be described in
terms of a $2 \times 2$ Green's function, which obeys the Gorkov's
equations. These equations can be simplified when the interesting
scale in the problem being considered is much larger than the
Fermi wavelength $\lambda_F$\cite{E68}. In the quasiclassical
approximation, the oscillations of the Green's function, on a
scale of $\lambda_F$, are averaged\cite{Rammer86}. In the dirty
limit the Green's function is almost isotropic and an expansion in
spherical harmonics keeping only the $L=0$ term can be made. The
Green's functions are obtained from Usadel
equations.\cite{U70,Belzig99b} With these assumptions the Greens's
function can be parameterized in terms of two position and energy
dependent complex angle variables, $\theta (\vec{r},E)$ and
$\Phi(\vec{r},E)$\cite{Golubov88,Stoof96,Belzig99b}:

\begin{equation}
\hat{g} ( \vec{r} , E )\equiv \left( \begin{array}{cc} g & f \\
f^\dagger & g
  \end{array} \right )
\equiv \left( \begin{array}{cc} \cos [ \theta ( \vec{r} , E ) ]
&\sin [ \theta ( \vec{r} , E ) ] e^{i \Phi ( \vec{r} , E )} \\
\sin [ \theta ( \vec{r} , E ) ] e^{- i \Phi ( \vec{r} , E )} &\cos
[ \theta ( \vec{r} , E ) ] \end{array} \right) \label{param}
\end{equation}

\noindent $g$ being the ordinary propagator, $f$ the anomalous Green's
function and $f^\dagger$ its time reverse. Superconductivity is
characterized by a non-vanishing $f$, which gives the probability
amplitude for the destruction of a Cooper pair. With this
parametrization, the superconducting order parameter
$\Delta(\vec{r})$, which has to be determined selfconsistently,
and the density of states $N(\vec{r},E)$ are given by:

\begin{eqnarray}
\Delta(\vec{r})&= &N_0 V\int_0^{\omega_D} dE \tanh\left (
\frac{E}{K_B T}\right ) {\rm Im}
\left [ \sin \theta(\vec{r},E) \right ] \nonumber \\
N(\vec{r},E) &= &N_0 {\rm Re}\left [\cos \theta(\vec{r},E)\right ]
\end{eqnarray}

\noindent with $N_0$ the normal density of states, and $\omega_D$ the
Debye frequency. If there is no current $\Phi (\vec{r},E)$ is a
constant equal to the phase of the order parameter and we can
drop it in the following.

Let us consider an axially symmetric superconductor in a magnetic
field $H$ parallel to its axis. The superconductor is assumed to
be thin enough to neglect screening due to superconducting
currents, and the vector potential $\vec{A}$ is given by
$\vec{A}=\frac{1}{2}H r \vec{e}_\phi$. Usadel's equation can be
written as:

\begin{equation}
\frac{D}{2} \nabla^2 \theta + \left [ i E - \frac{1}{2 \tau_{in}}
\right ] \sin \theta + | \Delta | \cos \theta - \left [ \frac{1}{2
\tau_{pb}} + 2 e^2 D
  |  \vec{A} |^2 \right ] \cos \theta \sin \theta =0
\label{eq_Usadel}
\end{equation}

\noindent where $\tau_{in}$ and $\tau_{pb}$ are the inelastic and pair
breaking scattering times\cite{M66} and $D=\frac{1}{3}\ell v_F$
the diffusion coefficient. When the superconductor is thinner than
the coherence length, the radial dependence of the quantities of
interest can be neglected, and $\vec{A}^2$ replaced by its average
$\langle A^2(z) \rangle =H^2 R^2(z)/12$, where $z$ is the distance
to $\vec{r}=0$ along the superconductor, i.e.
parallel to H. The pair breaking effect of the magnetic field is
the given by:

\begin{equation}
\Gamma^H_0(z)=e^2 DH^2 R^2(z)/6
\label{eq_PB}
\end{equation}

Note that $\Gamma^H_0$ becomes strongly position dependent and
increases, for a fixed field $H$, with the square of the radius of
the sample $R(z)$.

\begin{figure}[p]
\begin{center}
\resizebox{0.7\columnwidth}{!}{
\includegraphics[width=10cm]{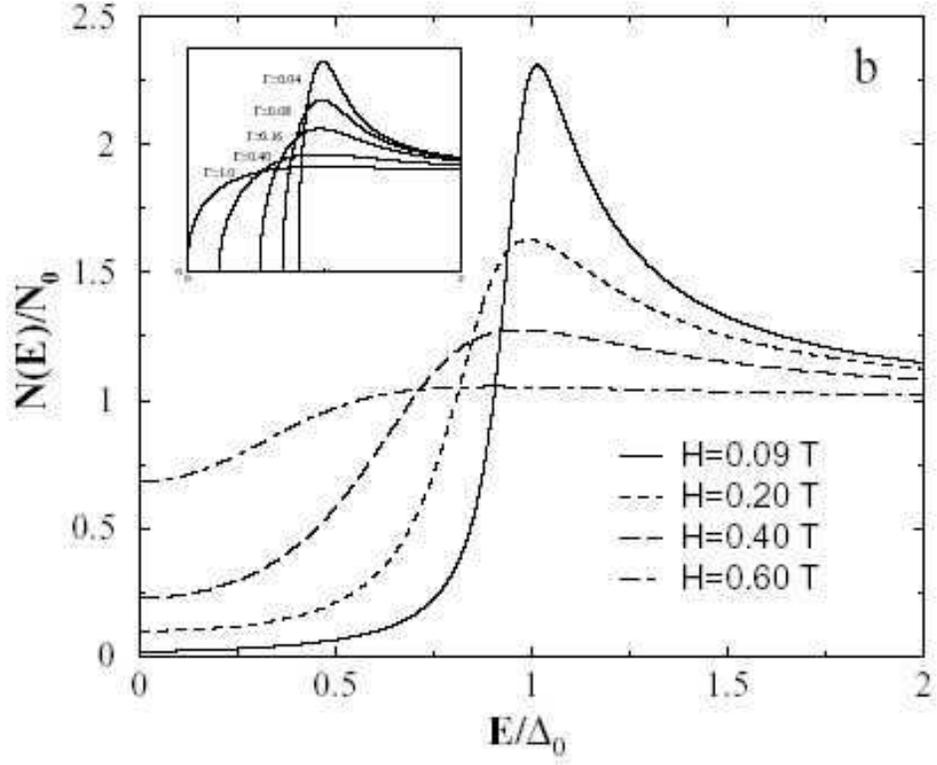}}
\caption{The density of states at the neck calculated using VRPB
model, and corresponding to $L=850\AA, \alpha=45^o$ (see inset of
Fig.\ref{fig_gap}) is shown as a function of the magnetic field.
The coherence length is reduced to $\xi=325\AA$ to model the
changes in $\ell$ at the nanobridge. In the inset the density of
states calculated using a pair breaking model, is represented for
different values of $\Gamma$. Note the absence of low energy
excitations in a large range of $\Gamma$, a point which is not
reproduced in the experiments, as discussed further on. From
\protect\cite{Suderow00c,B00}.} \label{densidad}
\end{center}
\end{figure}

In uniform wires, where $R$ is constant, the pair-breaking term
due to the field is also constant in all the wire and $\nabla
\theta =0$. Neglecting inelastic scattering processes and pair
breaking effects other than those due to the applied field, the
following equation is obtained:

\begin{equation}
\frac{E}{|\Delta|}=u\left(1- \frac{\Gamma}{\sqrt{u^2 -1}}\right )
\end{equation}

\noindent where $\Gamma=\Gamma_0^H/|\Delta|$ and the parameter $u$ is
defined by $\cos \theta=\frac{u}{\sqrt{u^2 -1}}$ and $\sin
\theta=\frac{-i}{\sqrt{u^2 -1}}$. In this way, the description for
uniform superconductors with pair breaking effects is
recovered\cite{M66}. The magnetic field penetrates in the
superconducting region, inducing pair breaking effects and
reducing the superconducting gap. The density of states, shown in
the inset of Fig. \ref{densidad}, remains gapped up to fields very
close to the critical one ($H_{gap}\sim 0.9 H_c^{wire}$, with
$H_c^{wire} \sim \left ( \frac{3}{e^2 D} \right )^{1/2}
\frac{1}{R}$).

\begin{figure}[p]
\begin{center}
\resizebox{0.7\columnwidth}{!}{
\includegraphics[width=10cm]{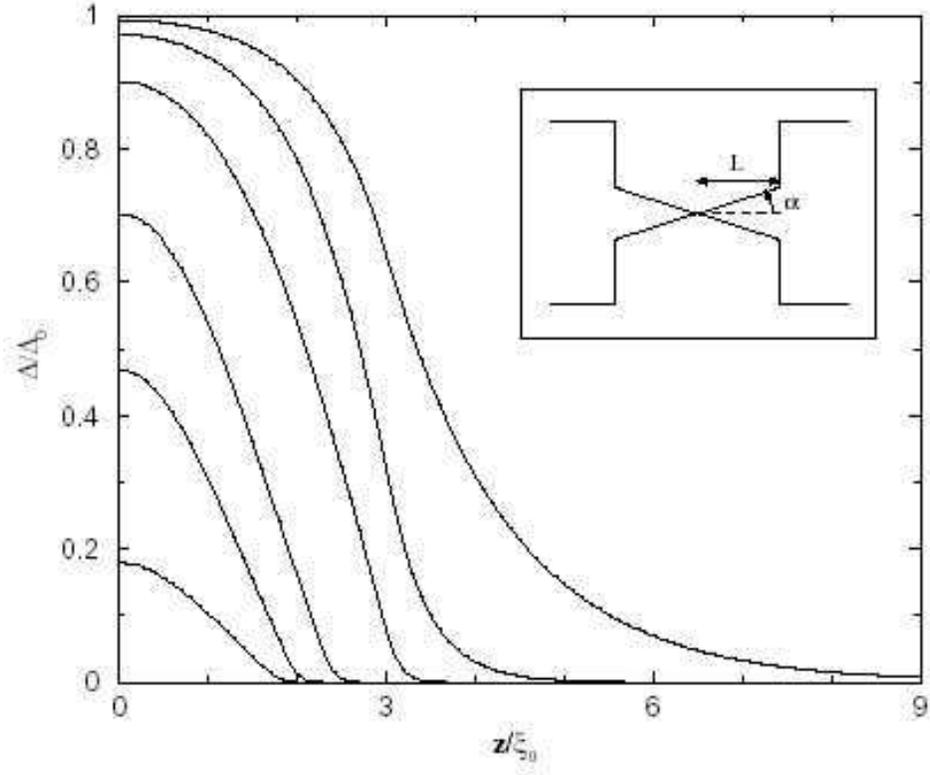}}
\caption{The inset shows the geometry used in the variable radius
description (VRPB). The main figure shows the dependence of the
superconducting order parameter, within VRPB, as a function of the
distance from the center of the nanobridge, for different applied
fields (from top to bottom, 0.09, 0.2, 0.4 and 0.6 T), using the same geometry
as in the previous figure. The neck is
joined at the electrodes at $z/\xi=2.6$. From
\protect\cite{Suderow00c,M66}} \label{fig_gap}
\end{center}\end{figure}

In cone-like nanobridges $R (z)$ and $\Gamma^H_0(z)$ smoothly
increase with the distance $z$ with respect to the center of the
structure at a given magnetic field $H$. The superconducting order
parameter will be eventually suppressed at a certain distance,
creating an N-S-N structure, in which only the central part of the
nanobridge remains in the superconducting state. From
eq. (\ref{eq_Usadel}), the magnetic field, temperature and position
dependence of the superconducting order parameter and density of
states is obtained by introducing a $R(z)$ function which
reproduces the geometry of typical nanobridges.
$R$ becomes a simple linear function of $z$ 
if the nanobridge is modelled by two
truncated cones each of length $L$, with an opening angle
$\alpha$, joined by their vertices and attached to bulk
electrodes, as shown in the inset of Fig.\ref{fig_gap}
\cite{Suderow00c,B00}. 
Fig.\ref{fig_gap} shows the
superconducting order parameter as a function of the distance for
different applied fields and typical $\alpha$ and $L$. There is a
smooth transition to the normal state as the radius of the
nanobridge increases. The magnetic field dependence of the density
of states at the center of the nanobridge, calculated using the
same $R(z)$ as in fig.\ref{fig_gap}, is shown in
fig.\ref{densidad}. A large amount of low energy excitations,
induced by the proximity effect from the normal parts of the
nanobridge (fig.\ref{fig_gap}), is found in the whole field range.
In fact, the superconducting gap is lost already at fields very
close to the bulk H$_c$ (0.08T), which contrasts the case of a uniform
wire (inset of fig.\ref{densidad}). As shown below, these
differences can be addressed experimentally.

\subsubsection{Ginzburg-Landau approach.}

The Ginzburg-Landau (GL) approach \cite{GL50} has been extensively
applied to the study of the distribution of the magnetic field and
the superconducting phase in small superconducting
systems\cite{Metal95,Getal97,Getal98}. A very good account of the
geometrical distribution of the order parameter is obtained,
although the details of the density of states are more difficult
to address. Its applicability is, in principle, restricted to
temperatures close to $T_c$, where it becomes equivalent to
Usadel's formalism \cite{M64}.

The theoretical analysis makes use of the similarities between the
Ginzburg-Landau equations for planar superconductors in a magnetic
field and the Schr\"odinger equation for a particle in a
field\cite{P98}. The predicted magnetic structure is very
rich\cite{SP99,P00}, and it has been compared with
experiments\cite{Getal00,Misko01}.
Characteristic vortex configurations observed in refs.
\cite{Getal00,Misko01} originate from the competition between 
the vortex-vortex interaction
and the vortex-sample edge interaction \cite{Bapet}. 

\begin{figure}[p]
\resizebox{0.75\columnwidth}{!}{
\includegraphics[width=9.5cm]{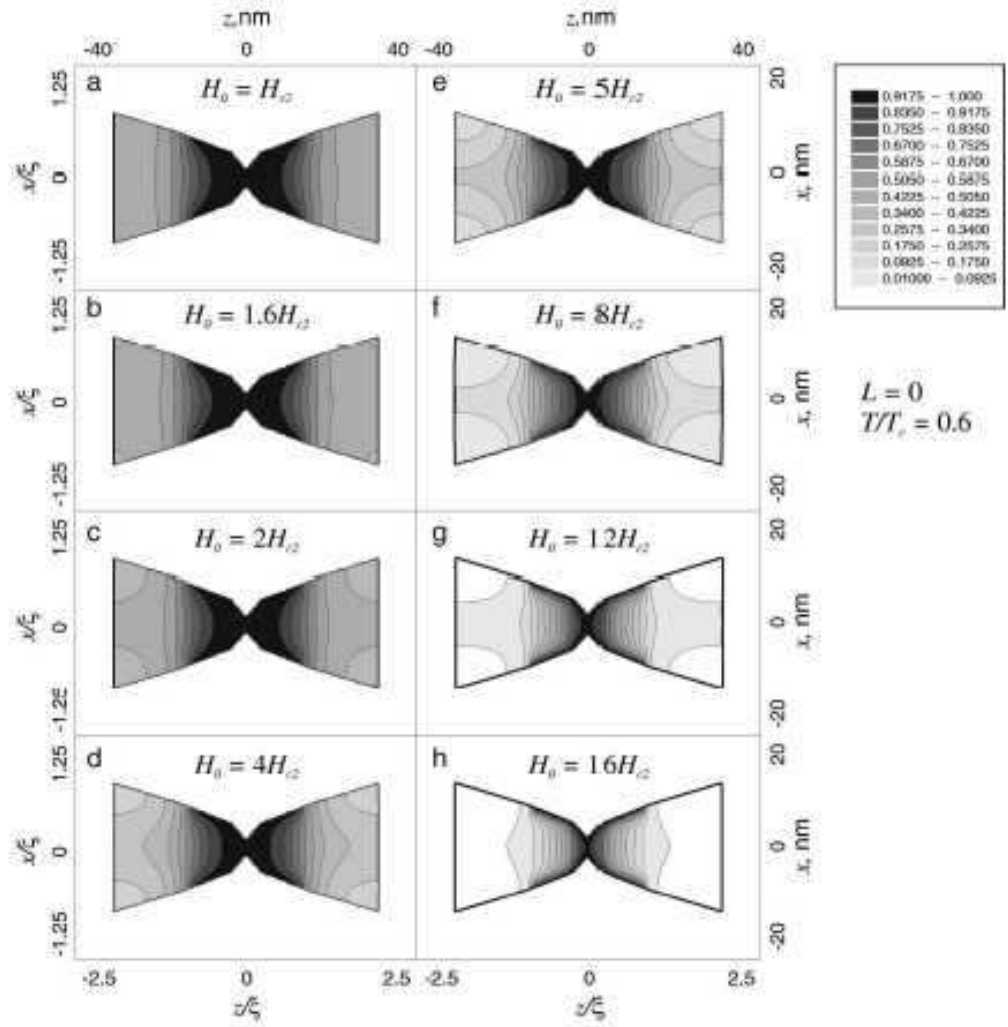}}
\caption{Distribution of the superconducting order parameter in a
nanostructure at different magnetic fields (at T/T$_c$=0.6)
(from\protect{\cite{MFD01}}). Superconductivity survives
at the central part of the neck up to fields much higher than the
bulk critical one.} \label{GL}
\end{figure}

GL equations have been also applied to study the distribution of
the superconducting phase in an isolated (non-connected to large
electrodes) cone-like geometry\cite{MFD01,MFD02}, which precisely
models the nanobridge created in \cite{Petal98} (see Fig.\ref{GL}). 
The 3D equations
have been solved, relaxing the assumption of an order parameter
constant in the radial direction. These results confirm the
applicability of the VRPB model. Already at low magnetic fields,
it is found that the parts of the nanobridge with larger radius
transit to the resistive state. At the highest magnetic fields,
the superconducting phase is concentrated near the neck of the
bridge, and very interesting, ring-shape areas, indicating states
with finite vorticity, appear in a short range of fields and at
low temperatures. These calculations have raised the question, not
yet addressed in published experiments, of the possible
confinement of vortices within the smallest possible
superconducting structure. The fundamental properties of these
vortex states represent, in our opinion, an interesting field for
future studies.

The modification of the critical field of a thin wire by the
presence of a current has been also calculated within the GL
approach. The critical current, in the absence of the field, is:

\begin{equation}
I_c ( T ) = \frac{H_c ( T ) R^2 c}{3 \sqrt{6} \lambda ( T )}
\label{Ic}
\end{equation}

In terms of this quantity, the critical field in the presence of a
current, and the critical current as function of field can be
written as\cite{B00}:
\begin{eqnarray}
H_c ( T , R , I ) &= &H_{wire} ( R , T ) \sqrt{1 - \left[
\frac{I}{I_c ( T )} \right]^{2/3}} \nonumber \\
I_c ( T , R , H ) &= &I_c ( R , T ) \left\{ 1 - \left[
\frac{H}{H_{wire} ( R , T )} \right]^2 \right\}^{3/2}
\label{Hc_Ic}
\end{eqnarray}

\noindent with
\begin{equation}
H_{wire}(T)=\sqrt{32}H_c\frac{\lambda(T)}{R}
\end{equation}

An extensive analysis of the functions $H_c$ and $I_c$ can be
found in\cite{B00}. It is found that a magnetic field above the
bulk critical value reduces the critical current. This reduction
is more important for wires with larger radius R.

\subsection{Experiments}

Previously we have discussed the experimental situation at zero
magnetic field in the different conduction regimes, tunnelling,
atomic size contact and weak link regimes. Experiments under
magnetic fields confirm the applicability of the VRPB and GL
models and show up new effects, associated with the presence of a
N-S interface near the center of the nanobridge.

\subsubsection{Tunnelling regime.}

The tunnelling conductance in the superconducting state has been
measured following the rupture of narrow and symmetric nanobridges
of Pb\cite{Suderow02,REPJ04,Suderow00d}. In these tunnel
junctions, superconducting correlations have been observed up to
magnetic fields as high as 2T, i.e. at fields 25 times higher than
the zero temperature critical field of bulk samples (0.08T). In
all cases, a large amount of low energy excitations is observed,
which can only be explained taking properly into account the
geometry of this structure, as in the VRPB model.

\begin{figure}[p]
\begin{center}
\resizebox{0.7\columnwidth}{!}{
\includegraphics[width=10cm]{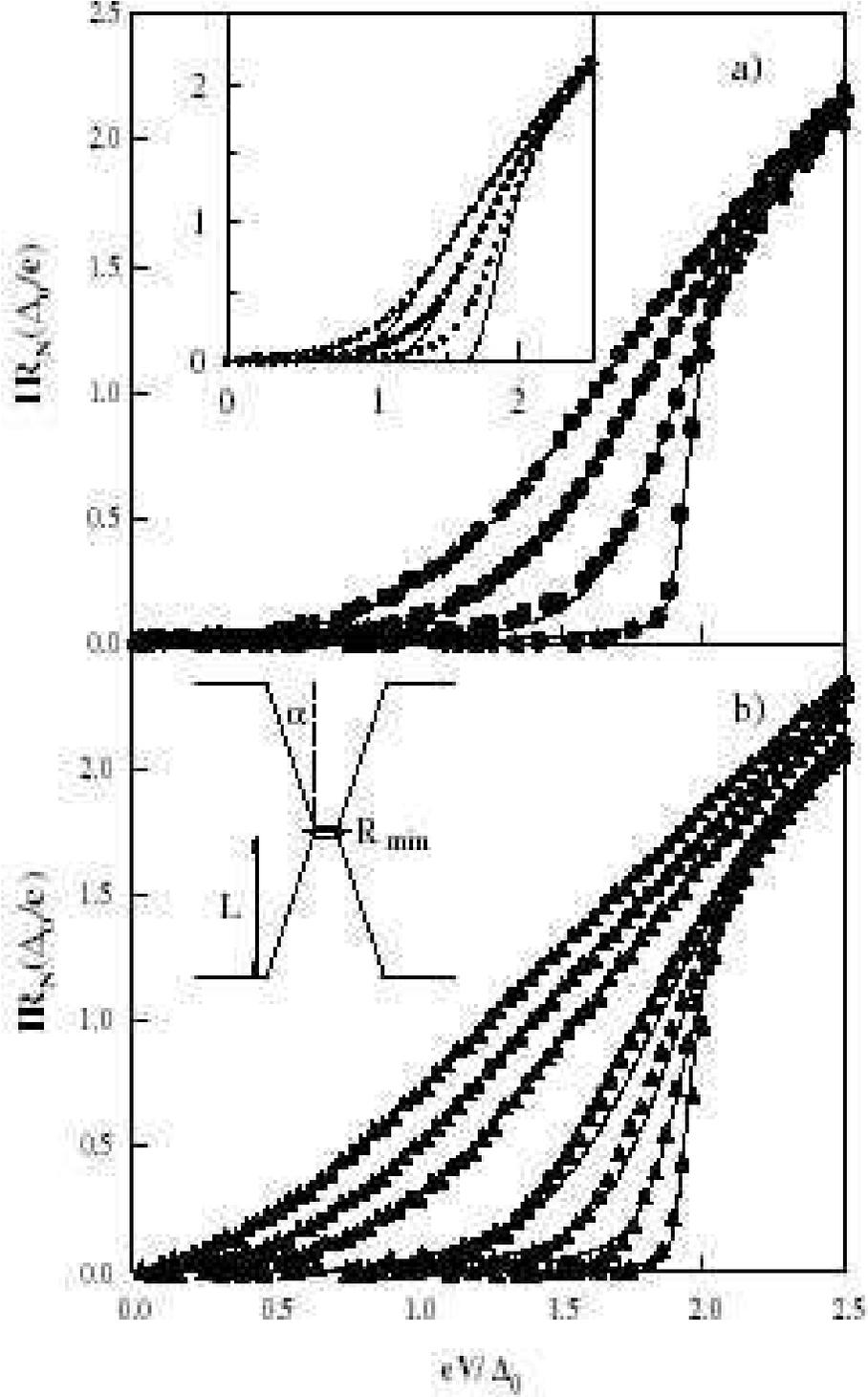}}
\caption{Current (symbols) as a function of bias voltage at
$T=0.4$ K (a), from bottom to top H = 0, 0.13, 0.18, 0.23 T, and
$T=1.5$ K (b) from bottom to top H = 0, 0.17, 0.34, 0.5, 0.84, 1.01,
1.18 T, for two characteristic nanobridges with the magnetic field
applied parallel to the bridge in (a) and perpendicular in (b).
Solid lines correspond to the fittings obtained within a variable
radius geometry (VRPB, see inset in b. and fig.\ref{fig_gap}).
Inset in a. shows calculations (lines) together with experimental
curves (points) using a single pair breaking parameter (inset in
fig.\ref{densidad}, from bottom to top:$\Gamma_0=0.04,0.13,0.21$).
This model does not reproduce the large measured current at low
voltages. From \protect\cite{Suderow02}.} \label{fig:Figtunel}
\end{center}
\end{figure}

I-V curves calculated with the constant pair breaking
model\cite{M66} do not reproduce the large current found in the
low voltage part of the measured I-V curves, as shown in the inset
of Fig.\ref{fig:Figtunel}a. By contrast, the VRPB model, using
parameters consistent with the geometry of typical nanobridges,
leads to calculated I-V curves, which do indeed follow the
experiment, as shown in Fig.\ref{fig:Figtunel}a) and b) for two
different samples. Note that the same geometry is used to fit the
whole field range. The large amount of current found in this range
is the result of the excitations induced by the proximity effect
from the parts of the nanobridge with largest radius, which
transit to the normal state at smaller fields than the central
part of the nanobridge(Fig.\ref{fig_gap}).  The destruction of
superconducting correlations is a combined effect of the proximity
from the perfectly connected normal parts around the
superconducting neck, and the pair breaking effect of the magnetic
field. These results show, quite unambiguously, that the magnetic
field restricts the volume of the superconducting phase to its
minimal size.

\begin{figure}[p]
\begin{center}
\resizebox{0.7\columnwidth}{!}{
\includegraphics[width=8.5cm,clip]{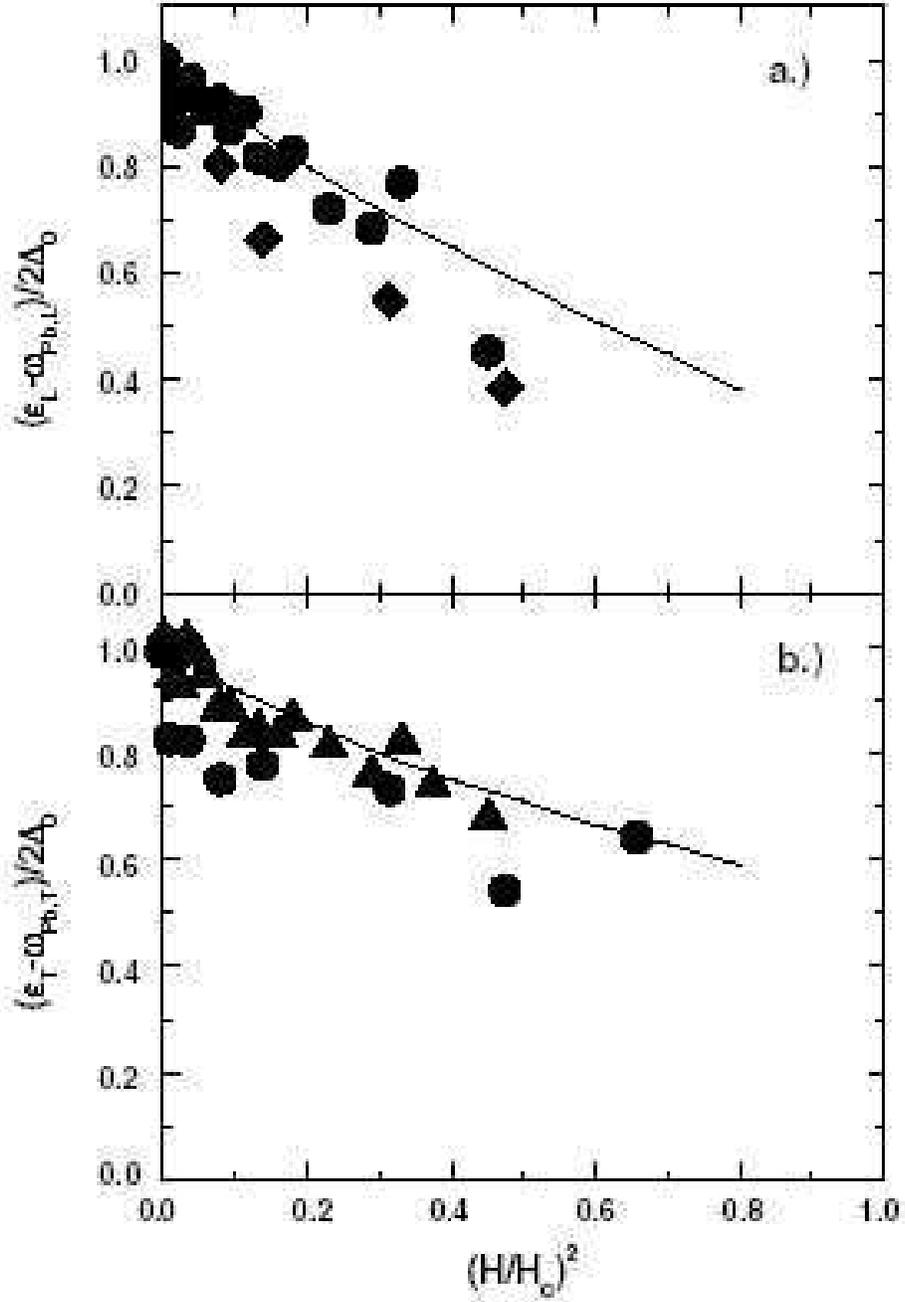}}
\caption{The strong coupling features corresponding to the
longitudinal (a) and transversal (b) phonon modes ($\omega _{L,T}$) is shown as a
function of the magnetic field. The data are extracted directly
from the position of the corresponding features in the tunneling
density of states ($\epsilon _{L,T}$), and plotted normalized to the superconducting
gap at each field. The theory developed by \protect\cite{Daams84}
(lines) explains the data. From \protect\cite{Suderow02}.}
\label{FigPhononH} \end{center}
\end{figure}

The features in the density of states corresponding to the phonon
structure (see Fig.\ref{Pbtunel1}) have been followed as a
function of the magnetic field in Ref.\cite{Suderow02}, giving the
unique possibility to study the pair formation when, gradually,
superconductivity is confined to the smallest length scales. In
Fig.\ref{FigPhononH} the variation of the characteristic phonon
modes is shown as a function of the magnetic field. To normalize
the position in energy of the phonon features, following previous
experiments and calculations for thin films in
Refs.\cite{Rasing81,Daams84}, the phonon frequencies
$\omega_{L,T}$ are subtracted from the voltage position of the
features corresponding to the transverse and longitudinal phonons
$\epsilon_{L,T}$, and divided by twice the zero field
superconducting gap $2\Delta_0$. The data (fig.\ref{FigPhononH})
follow well the calculations of \cite{Daams84}, intended to
explain the experiments in thin films \cite{Rasing81}. The
confinement of the superconducting correlations to its minimal
size by increasing the magnetic field occurs without specially
marked changes in the way the Cooper pairs are formed. The perfect
connection between the smallest superconducting part of the
nanobridge and the large bulk in the normal state, 
is the key element to
understand the superconducting state under magnetic fields in
these nanostructures.

\begin{figure}[p]
\begin{center}
\resizebox{0.7\columnwidth}{!}{
\includegraphics[width=10cm]{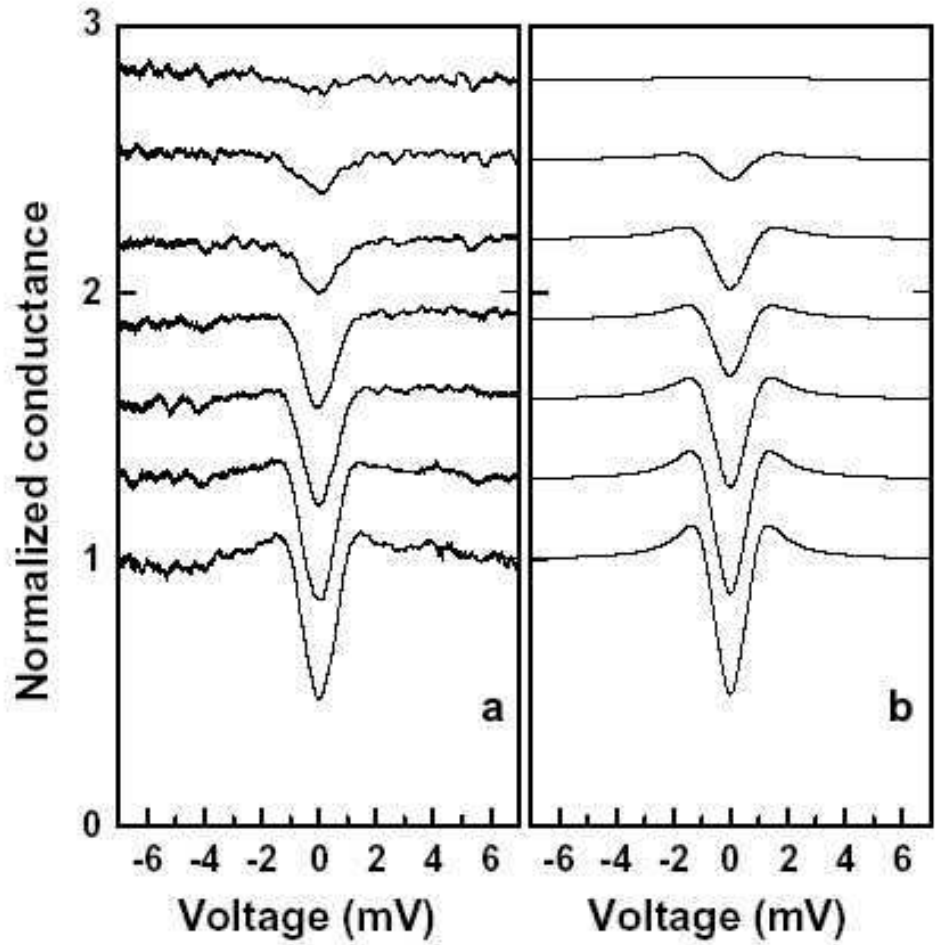}}
\caption{After the creation of the nanostructure, if the tip is
transported to a region of the sample which is flat, curves
characteristic of N-S junctions are obtained, with the
superconducting density of states showing a large amount of low
energy excitations (experiment, a), explained within the variable
geometry model of Ref.\protect\cite{Suderow00c} (calculated curves
in b). 
From bottom to top, $T = 1.0, 1.5, 2.0, 2.5, 3.0, 3.5, 4.0 K $ }\label{Fig:fp2t}
\end{center}
\end{figure}

Somewhat different experiments have been done by transporting one
half of the broken nanobridge, i.e. a nanotip (see Fig.\ref{fig:cartoon}), to
a more flat region onto the Pb sample\cite{REPJ04}. In that case,
the flat region of the sample is in the normal state, while the
tip remains superconducting under field, resulting in
characteristic N-S tunnelling curves, which can only be
satisfactorily explained within the VRPB model. In
Fig.\ref{Fig:fp2t} representative series of data taken as a
function of temperature (a) and its corresponding theoretical
calculations (b) are shown. Again, these results evidence the
strong reduction of the size of the superconducting part under
magnetic fields.

\subsubsection{Atomic size contact regime.}

The multiple Andreev reflection pattern has been followed as a
function of the magnetic field in many single atom point contacts
made in nanobridges with different geometries
\cite{B00,Suderow00c,Suderow00d}. The atomic arrangements around
the contacting atom are easily changed without destroying the
overall shape of the nanobridge, so that a large series of
contacts can be studied at a given magnetic field. The subharmonic
gap structure (SGS) is gradually smeared out by the magnetic
field.

\begin{figure}[p]
\begin{center}
\resizebox{0.7\columnwidth}{!}{
\includegraphics[width=10cm]{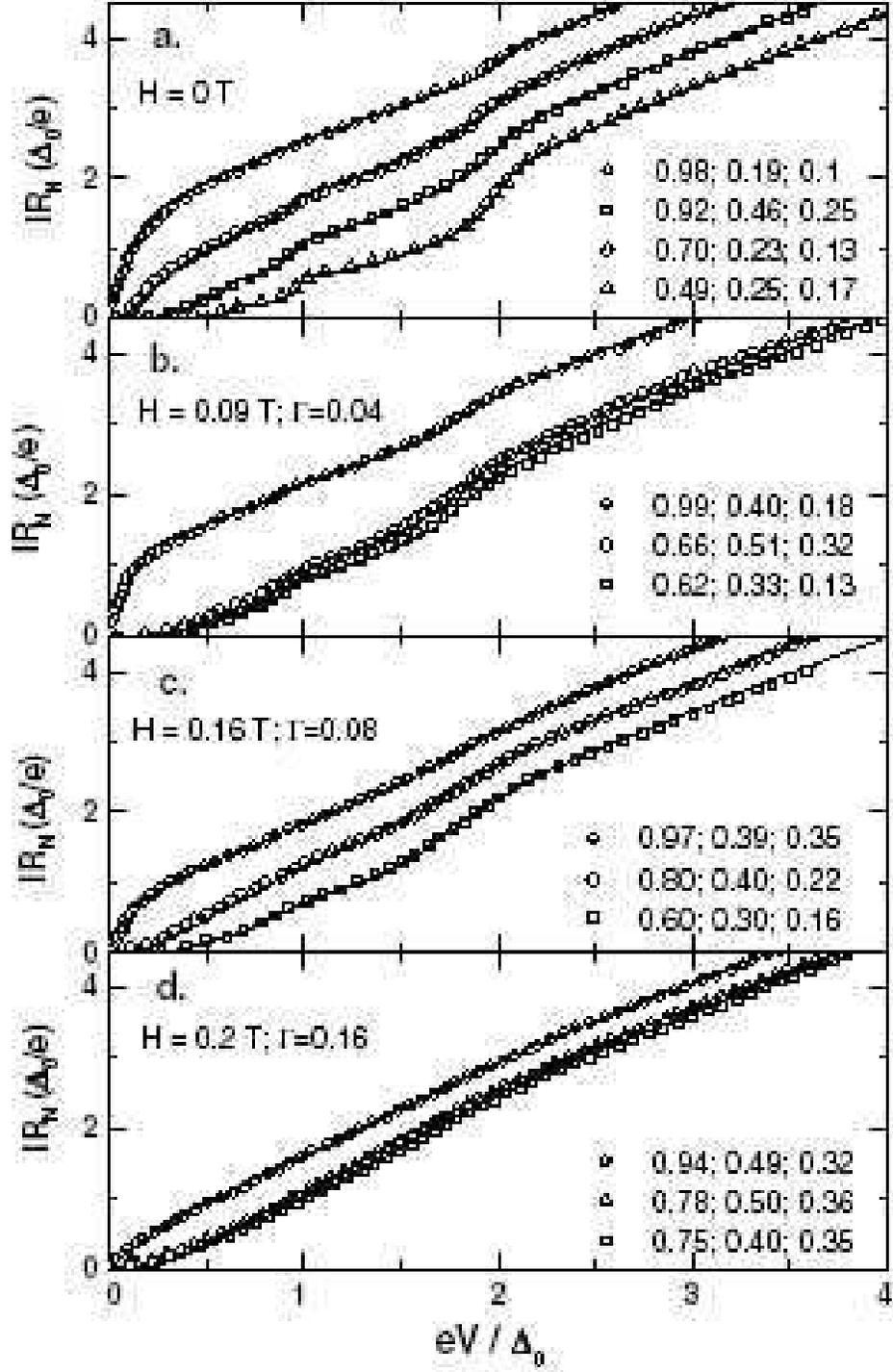}}
\caption{Comparison of theoretical (lines) and experimental
(symbols) single-atom contact I-V curves. Theoretical curves
correspond to the constant pair breaking model. Note that the
expected field dependence of $\Gamma \propto H^2$
(equ.\protect\ref{eq_PB}) within this model is not found when
trying to reproduce the experiment.
The parameters in the lower right corner of each figure are the
transmissions through the different channels used to fit the experimental
data. Each line of numbers corresponds to one curve, from top to bottom.
$\Gamma $ is the pair-breaking parameter defined in the text.} \label{contacto1}
\end{center}
\end{figure}

\begin{figure}[p]
\begin{center}
\resizebox{0.7\columnwidth}{!}{
\includegraphics[width=10cm]{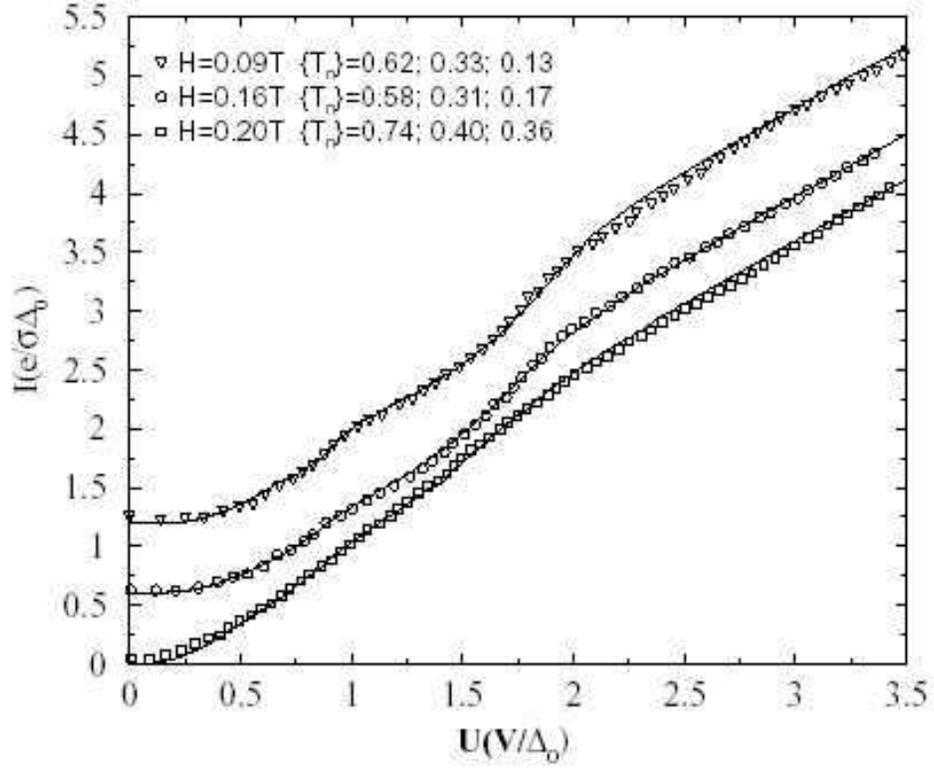}}
\caption{Comparison of several theoretical (lines) and
experimental (points) atomic size contact I-V curves (data are
shifted for clarity). Theoretical curves, calculated using 
the VRPB model, correspond to a geometry
with $L=850$\AA, $\alpha = 45^o$ and $\xi=325$\AA.}
\label{contacto2}
\end{center}
\end{figure}

I-V curves made under magnetic fields in some nanobridges have
been reproduced using a uniform pair breaking parameter $\Gamma$
and the typical distribution of conduction channels found in Pb
(fig.\ref{contacto1}). However, the magnetic field dependence of
$\Gamma$ was found to be inconsistent with formula (\ref{eq_PB}),
 making impossible to give a physical meaning to the
values of $\Gamma$ used within this description\cite{Suderow00c}.
Moreover, subsequent work \cite{B00} has found significant
discrepancies in situations where the N-S interface moves close to
the contact, as e.g. at fields close to the complete destruction
of superconductivity. The VRPB model, by contrast, gives an adequate
fit to the experiment in every case, as shown in a representative
example of curves in Fig.\ref{contacto2}. The same geometry,
compatible with the one determined experimentally, is used for a
given nanobridge in the whole magnetic field
range\cite{Suderow00c}.

A remarkable conclusion of these experiments is that the magnetic
field does not change the essential properties of the atomic size
contact, in the sense that the same distribution of conduction
channels is needed to fit the experiments. When the field is
applied no conduction channel is closed nor a new channel
opens\cite{Suderow00c}. In fact, the flux going through the
contact, even at fields of several T, is much too small to produce
changes in the electronic transport in the neck or in the orbital
structure of the contacting atom.

Other experiments made with the break junction technique in Al
also show how atomic size contact curves are smeared by the
application of a magnetic field\cite{Scheer00}. In those
experiments, one contact was followed as a function of the
magnetic field to maintain the distribution of channels constant
in the whole field sweep. Their results were interpreted using the pair
breaking model to account
for the influence of the magnetic field. However,
superconductivity was lost at the bulk critical field of Al
(9.9mT). This technique does not produce nanobridges, so that the
region surrounding the contact has the same magnetic response as
the bulk, which is in the Meissner state. The
magnetic field distribution around the contact should be complex
due to demagnetization effects.

The fabrication of a nanobridge around the contact, and the
concomitant observation of superconductivity above the bulk
critical field, is necessary to get an accurate control over the
magnetic response of the system.

\subsubsection{Weak link regime.}

As discussed above (section III.B.3), cone-like structures which
are connected through a large contact of several nanometer in
radius show considerable local overheating
effects(Fig.\ref{JJheat}). As a consequence, a bump, indicative of
the loss of the excess current (formula (\ref{exceso})), appears in
the differential conductance. Under magnetic fields, the position
of the bump moves to smaller voltages, due to the decrease of the
superconducting order parameter.

Interesting situations have been reported in which the conductance
shows two bumps whose voltage positions have different magnetic
field behaviors. These results have been interpreted as being
characteristic of asymmetric nanobridges, in which each half of
the cone-like structure has a different opening angle, and
therefore also a different power dissipation rate\cite{Rodrigo03}.
The half with a smaller opening angle is more easily heated than
the half with a larger opening angle. However, the former has a
higher critical field than the latter. Correspondingly, while the
position of the feature corresponding to the first loss of the
excess current (the side with the smaller opening angle becoming
normal) remains almost constant, the voltage at which the second
loss takes place (the side with larger opening angle) varies
strongly with the applied field. These experiments show that it is
possible to obtain relevant information about the nanobridge by a
careful analysis of the observed I-V curves in the weak link
regime.

\begin{figure}[p]
\begin{center}
\resizebox{0.7\columnwidth}{!}{
\includegraphics[width=8.5cm,clip]{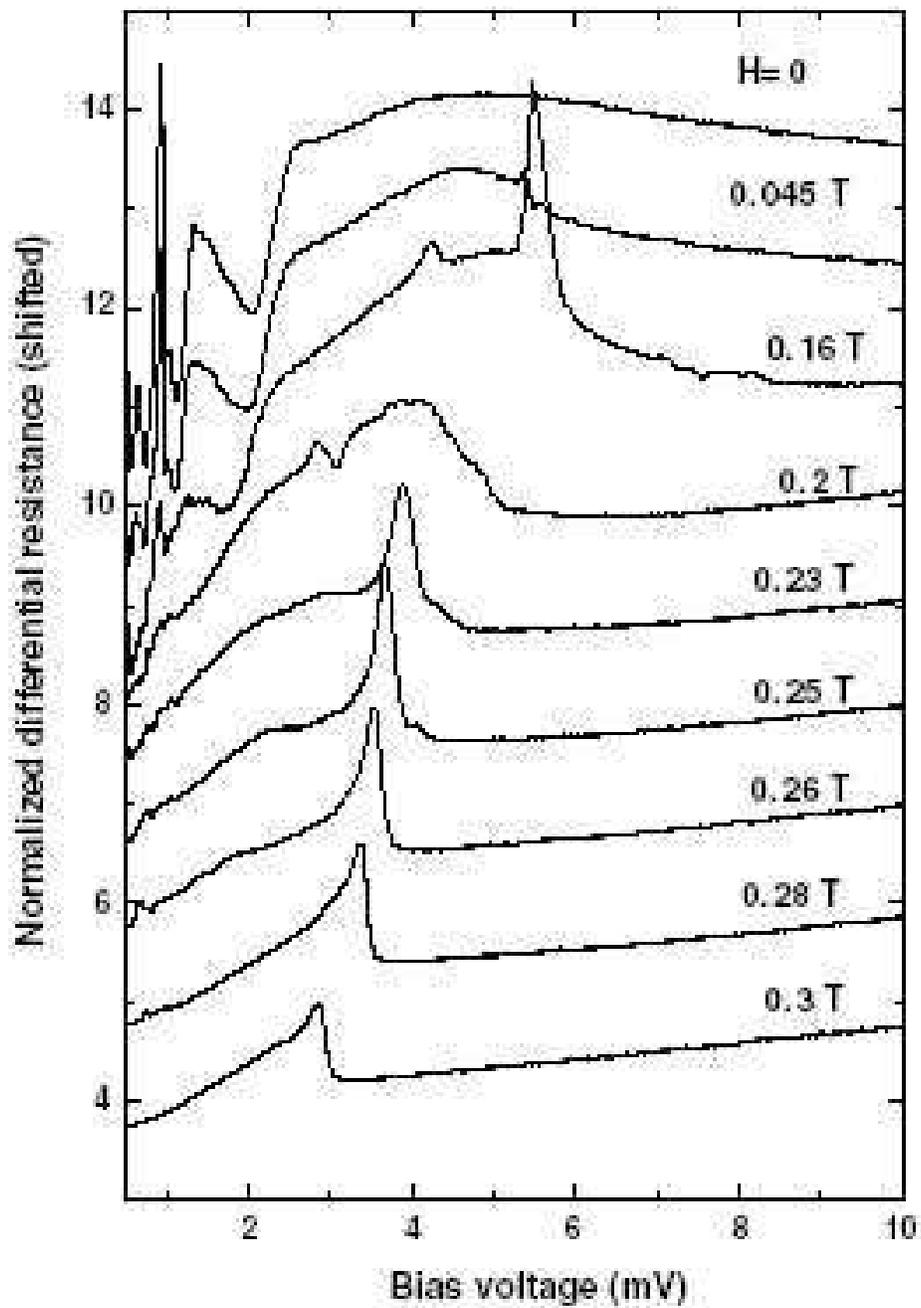}}
\caption{Nanobridges showing heating effects, also show a striking
behavior when the magnetic field is applied. Characteristic peaks
appear in the differential resistance, indicative of the
nucleation of phase slip centers. Above 0.23T, a single peak
structure appears in the conductance, which has been related to
non-equilibrium effects. From \protect\cite{Suderow00b}.}
\label{overallPS} \end{center}
\end{figure}

On the other hand, in sufficiently long and narrow nanobridges
with a symmetric geometry a striking phenomenology has been found
with two well defined regimes as a function of the magnetic
field\cite{Suderow00b}. A series of peaks appears in the
differential resistance below a magnetic field which corresponds
to about half of the field for the complete destruction of
superconductivity (0.2 T in Fig.\ref{overallPS})
\cite{Suderow00b}. These peaks are located at different voltages,
but always well above the superconducting gap, and have been
associated to the appearance of resistive centers in the
nanobridge, in close analogy to the phase slip centers observed in
thin wires\cite{T96}. At higher fields (above 0.23T in
Fig.\ref{overallPS}), a new regime sets in. The differential
resistance steeply increases with the voltage, and abruptly drops
to its normal state value when the local critical temperature is
reached. 
This results in a single sharp peak, instead of the
smooth structure observed at zero field, which disappears abruptly
at the critical temperature, and is always observed 
when the N-S interface moves close to the central part
of the nanobridge, as for instance near T$_c$\cite{Suderow00b,Rodrigo03}.
Therefore, it has been
associated to the establishment of a region within the
superconducting nanobridge, where a finite voltage drops. This
voltage has been attributed to a non-equilibrium situation created
by the conversion of normal current into a supercurrent, as
expected near N-S interfaces\cite{Schmid79}. Several experiments
and theoretical calculations have addressed the observation of
such non-equilibrium voltages in thin films, nanolithographed
structures, or nanoscopic wires
\cite{T96,Clarke79,Schmid79,Schmid80,Baratoff82,Escudero84,Mos94,Boogaard03,
Michotte04}.
The nanobridges created with the STM have two remarkable
differences with respect to those structures. First, the N-S interface is
formed in a natural way within the same material, eliminating
possible interface problems, which may appear in evaporated
structures. Second, nanobridges created by the STM also have much smaller
lateral dimensions enhancing the observed non-equilibrium signals.

Finally, we discuss the Josephson effect, which has been treated
in the literature in the large-contact regime with more detail than in
the tunnelling and single atom atomic size contact regimes. In
Ref. \cite{Petal98}, the critical current has been followed as a
function of the size of the contact. The qualitative trend is
explained with the calculations using GL model. In
Ref. \cite{Rodrigo03}, the form of the I-V curve near zero bias is
studied. It is shown that at zero field the zero bias conductance
does not diverge, but remains finite between $0.999T_c$ and $T_c$,
an effect typically observed in weak links due to thermal
activated phase slip through the barrier\cite{T96}. However, this
temperature range strongly increases in nanobridges under magnetic
fields. In some samples, a finite zero bias conductance has been
observed between $0.3T_c$ and $T_c$. This is a striking result,
and it is unclear if it can solely be explained by the modification
of the superconducting order parameter in the VRPB
model\cite{Rodrigo03}.

\newpage

\newpage

\newpage

\newpage

\newpage

\newpage

\newpage

\newpage

\newpage

\newpage

\newpage

\newpage

\newpage

\newpage

\newpage

\section{Advances and future prospects}

The nanostructures discussed in this topical review open new interesting 
fields for basic research and applications. In the following, we will mention 
some lines in which theoretical and experimental advances 
could emerge in a near future.

\subsection{Theory and fundamental properties}

The properties of nanoscopic superconducting bridges in a magnetic
field can be reasonably described using adequate extensions of the
BCS theory with bulk values for the gap and other relevant
parameters. The electron-electron and electron-phonon interactions
should be modified near the contact. The structures discussed here
are very well suited to analyze these effects. It would be
interesting to estimate the influence of the geometry of the
contact on the pairing interactions.
Some calculations on the the properties of these nanobridges under 
magnetic field \cite{Misko01}  have raised the question, not
yet addressed in published experiments, of the possible
confinement of vortices within the smallest possible
superconducting structure. The fundamental properties of these
vortex states represent, in our opinion, an interesting field for
future studies.

The existence of a close contact between a region of the
superconducting phase and a bulky normal phase can allow to study
small differences between the two phases not included in the BCS
theory. An interesting effect is a possible charge transfer
between the two phases, derived theoretically from different
assumptions\cite{KK92,KF95,Letal04,H04,H04b}.

The electron-electron interactions in mesoscopic devices closely
connected to bulk electrodes is a topic of significant theoretical
interest\cite{N99,GZ01}, as these systems show Coulomb blockade
features similar to those found in weakly coupled
systems\cite{GS87,SET}. It will be interesting to know if this
enhancement of the repulsive interactions modify the
superconducting properties of the junction at very low
temperatures.

The systems studied here, as discussed earlier, allow us to study
superconductors where the gap is not constant throughout the Fermi
surface. In these materials, elastic scattering by lattice
imperfections is a source of pair breaking processes, playing a
similar role to scattering by magnetic impurities in ordinary
superconductors. The surface of the sample itself leads to
transitions between different bands, or different regions of the
Fermi surface\cite{BG01}, and it can change locally the
superconducting properties of materials where the gap is not
constant, as it is probably the case in MgB$_2$\cite{Martinez03}.
This effect is not yet studied in detail.

\subsection{On the use of superconducting tips}

The capability of the method described in section 2 to prepare
and characterize at low temperatures STM tips will open new possibilities.
Cleanness and atomic sharpness can be guaranteed and easily restored if 
it is needed during operation.
One of the applications recently pointed out is the use of well 
characterized superconducting tips to do scanning Josephson 
spectroscopy (SJS)\cite{Balatsky,Stip2}. 
This technique is a very promising way to obtain important information on 
the nature 
of the order parameter of several superconducting materials,
especially when a non-conventional behavior opens new challenges
\cite{INC99,Suderow04,Flouquet,TailleferRev,Davis}.
As noted in section 4, the knowledge achieved on the behaviour under magnetic fields of 
those superconducting tips is also important.
As it has been remarked, the electronic density of states at the tip can
be changed by modifying its shape, temperature and external magnetic field.
This leads to a series of situations, in which the superconducting
condensate is confined to a nanometric size region\cite{Rodrigo03},
 whose relevance goes
beyond the phenomena discussed in this Review.
In situations where
Zeeman effect is important, superconducting tips under magnetic
fields should show a spin split density of states
\cite{Mersevey94}, becoming a very sensitive probe of the local
spin polarization of the sample.
Samples with magnetic inhomogeneities such as vortices or ferromagnetic
domains will alterate the density of states of the tip while scanning
above them, providing a new nanoprobe within the family of the STM.

\subsection{Fluctuations and non-equilibrium effects}

During past years, several works have shown that in wires of
lateral dimensions even much larger than those of the nanobridges
discussed in this review, the resistivity does not go to zero when
cooling well below
T$_c$\cite{TinkhamQPS,TinkhamQPS2,Vodolazov03,Boogaard03,Michotte04}.
This is explained by the appearance of quantum phase slip centers,
i.e. phase slippage induced by quantum tunnelling through the
relevant energy barrier, and not by thermal activation. The
fundamental question about the way to include the influence of
dissipation \cite{Schmid83} in those systems has been the subject
of recent debate, and the resistance of the wires has been
advanced as the actual source of
dissipation\cite{TinkhamQPS,TinkhamQPS2}. The nanobridges subject
of this review are perfectly joined to the bulk, and represent an
interesting alternative way of confining superconductivity, which
seems, a priori, to lead to more stable properties at all
temperatures. The role of the dissipation should be associated to
the proximity to the bulk, instead on the resistance of the
structure, and it should be very strong in any case. The large
range of temperatures in which a finite zero bias conductance has
been observed under magnetic fields \cite{Rodrigo03}, contrasts
with the observations at zero field and shows that the confinement
of the condensate to the smallest length scales increases the
influence of fluctuations. The role of quantum fluctuations 
in explaining this behavior of nanobridges remains
an interesting question for future work.

\subsection{The ultimate nanostructure}

In a recent experiment G. Rubio Bollinger et al.\cite{Rubio03}
have used the STM to study the transport between two proximity
induced superconducting electrodes. Nanocontacts were produced
between two wires of Pb, in an original experimental arrangement
which permitted to study many different contacts.
The wires were arranged as a cross in such a way as to change
easily the position where contact is made by using an in-situ
working x-y table. The wires were made of bulk lead covered by a
thick layer of lead (approx. 900 nm), ensuring a clean surface,
and subsequently evaporating, in-situ on top of the Pb layer, a
thin layer of gold, approx. 28 nm in width. In order to minimize
inter-diffusion between both elements, the sample was in good
thermal contact with a sample holder refrigerated by liquid N$_2$.
Clean single atom point contacts, and atomic chains of gold
between the two wires could be routinely made with this technique.
In the superconducting phase, multiple Andreev reflection
processes occur, allowing for a precise determination of the
conducting channels across the single atom point contact. As
theoretically expected, in all cases, only a nearly completely
opened single channel has been found, not only in single atom
point contacts but also in atomic chains consisting of up to five
atoms arranged one after the other. To our knowledge the atomic
chain fabricated by these authors using a STM is possibly the
smallest weak link between two superconductors ever made. It will
be interesting to determine if the chain length affects or not the
coupling between condensates (Josephson effect).

\section{Acknowledgments}
We are grateful to our colleagues N. Agra\"\i t, J.P. Brison, P.C. Canfield, 
G. Crabtree, J.T. Devreese, J. Flouquet, V.M. Fomin, J.E. Hirsch, A. Levanyuk, 
K. Maki, V.V. Moshchalkov, O. Naaman, Y. Pogorelov, G. Rubio-Bollinger, 
G. Sch\"on and R. Villar for their help and stimulating discussions concerning 
this work at different stages.
We acknowledge  support from the European Science Foundation programme VORTEX,  
the MCyT, Spain (grants MAT-2001-1281-C02-0 and MAT2002-0495-C02-01),  
the Comunidad Aut\'onoma de Madrid, Spain (projects 07N/0039/2002 and 07N/0053/2002), 
the Swiss National Science Foundation and NCCR MANEP. The Laboratorio de Bajas 
Temperaturas is associated to the ICMM of the CSIC.

\bibliography{JGRrev1}

\end{document}